\documentclass[a4paper,fleqn,usenatbib]{mnras}

\usepackage{newtxtext,newtxmath}

\usepackage[T1]{fontenc}
\usepackage{ae,aecompl}

\usepackage{graphicx}	
\usepackage{amsmath}	

\defcitealias{clerc2016}{C16} 	 
\defcitealias{kirkpatrick2020}{K20}

\title[SPIDERS overview and spectroscopic data]{SPIDERS: overview of the X-ray galaxy cluster follow-up and the final spectroscopic data release}

\author[N. Clerc et al.]{
N. Clerc,$^{1}$\thanks{E-mail: nicolas.clerc@irap.omp.eu}
C.~C. Kirkpatrick,$^{2,3}$
A. Finoguenov,$^{2,4}$
R. Capasso,$^{5}$
J. Comparat,$^{4}$
\newauthor
S. Damsted,$^{2}$
K. Furnell,$^{6}$
A. E. Kukkola,$^{2}$
J. Ider Chitham,$^{4}$
A. Merloni,$^{4}$
M. Salvato,$^{4}$
\newauthor
A. Gueguen,$^{4}$
T. Dwelly,$^{4}$
C. Collins,$^{5}$
A. Saro,$^{7,8,9}$
G. Erfanianfar,$^{4}$
D.~P. Schneider,$^{10,11}$
\newauthor
J. Brownstein,$^{12}$
G.~A. Mamon,$^{13}$
N. Padilla,$^{14}$
E. Jullo,$^{15}$
D. Bizyaev$^{16}$
\\
$^{1}$IRAP, Universit{\'e} de Toulouse, CNRS, UPS, CNES, F-31028 Toulouse, France\\
$^{2}$Department of Physics, University of Helsinki, Gustaf H{\"a}llstr{\"o}min katu 2, FI-00014 Helsinki, Finland\\
$^{3}$Helsinki Institute of Physics, Gustaf H{\"a}llstr{\"o}min katu 2, FI-00014 Helsinki, Finland\\
$^{4}$Max-Planck-Institut f{\"u}r extraterrestrische Physik, Giessenbachstrasse, D-85748 Garching, Germany\\
$^{5}$The Oskar Klein Centre, Department of Physics, Stockholm University, Albanova University Center, SE 106 91 Stockholm, Sweden\\
$^{6}$Astrophyics Research Institute, Liverpool John Moores University, IC2, Liverpool Science Park, 146 Brownlow Hill, Liverpool L3 5RF, UK\\
$^{7}$Astronomy Unit, Department of Physics, University of Trieste, via Tiepolo 11, I-34131 Trieste, Italy\\
$^{8}$IFPU -- Institute for Fundamental Physics of the Universe, Via Beirut 2, I-34014 Trieste, Italy\\
$^{9}$INAF -- Osservatorio Astronomico di Trieste via G.B. Tiepolo 11, I-34143 Trieste, Italy\\
$^{10}$Department of Astronomy and Astrophysics, The Pennsylvania State University, University Park, PA 16802, USA\\
$^{11}$Institute for Gravitation and the Cosmos, The Pennsylvania State University, University Park, PA 16802, USA\\
$^{12}$Department of Physics and Astronomy, University of Utah, 115 S. 1400 E., Salt Lake City, UT 84112, USA\\
$^{13}$Institut d'Astrophysique de Paris (UMR 7095: CNRS \& Sorbonne Universit\'e), F-75014\\
$^{14}$Instituto de Astrof{\'i}sica, Pontificia Universidad Cat{\'o}lica de Chile, Av. Vicuna Mackenna 4860, 782-0436 Macul, Santiago, Chile\\
$^{15}$Aix Marseille Univ, CNRS, CNES, LAM, F-13388 Marseille, France\\
$^{16}$Apache Point Observatory, P.O. Box 59, Sunspot, NM 88349\\
}

\date{Accepted XXX. Received YYY; in original form ZZZ}

\pubyear{2019}

\begin{document}
\label{firstpage}
\pagerange{\pageref{firstpage}--\pageref{lastpage}}
\maketitle

\begin{abstract}
SPIDERS (The SPectroscopic IDentification of \emph{eROSITA} Sources) is a large spectroscopic programme for X-ray selected galaxy clusters as part of the Sloan Digital Sky Survey-IV (SDSS-IV).
We describe the final dataset in the context of SDSS Data Release 16 (DR16): the survey overall characteristics, final targeting strategies, achieved completeness and spectral quality, with special emphasis on its use as a galaxy cluster sample for cosmology applications.
SPIDERS now consists of about 27,000 new optical spectra of galaxies selected within 4,000 photometric red sequences, each associated with an X-ray source. The excellent spectrograph efficiency and a robust analysis pipeline yield a spectroscopic redshift measurement success rate exceeding 98\%, with a median velocity accuracy of $20$\,km\,s$^{-1}$ (at $z=0.2$).
Using the catalogue of 2,740 X-ray galaxy clusters confirmed with DR16 spectroscopy, we reveal the three-dimensional map of the galaxy cluster distribution in the observable Universe up to $z\sim0.6$. We highlight the homogeneity of the member galaxy spectra among distinct regions of the galaxy cluster phase space. Aided by accurate spectroscopic redshifts and by a model of the sample selection effects, we compute the galaxy cluster X-ray luminosity function and we present its lack of evolution up to $z=0.6$. Finally we discuss the prospects of forthcoming large multiplexed spectroscopic programmes dedicated to follow up the next generation of all-sky X-ray source catalogues.
\end{abstract}

\begin{keywords}
galaxies: clusters: general -- cosmology: observations -- X-rays: galaxies: clusters
\end{keywords}



\section{Introduction\label{sect:introduction}}

Studies of the distribution of galaxy clusters in the universe rely on large and well-selected samples of astrophysical objects tracing the most massive nodes of the cosmic web. Large amounts of hot gas trapped within their deep matter potential wells make clusters of galaxies luminous X-ray sources \citep[e.g.][]{jones1999}. They appear as diffuse extended sources \citep{abell1958, bahcall1993, bohringer2005, oguri2018, ridl2017, ricci2018, wen2018, andreon2019, gozaliasl2019}, with galaxy overdensities extending out to over 10\,Mpc \citep{trevisan2017}. Galaxies within halos may have their properties affected by encounters with the hot gas \citep[e.g.,][]{popesso2015, lotz2019, owers2019}.

Clusters of galaxies are rare objects, the more massive, the rarer. Their number density scales between $10^{-5}$ and $10^{-8}$\,Mpc$^{-3}$ for lower mass limits of $10^{14}$ and $10^{15}$\,$M_{\odot}$, respectively \citep[e.g.][]{vikhlinin2009}. X-ray surveys of the extragalactic sky fulfil the requirements of a large volume coverage, high completeness and high purity as required by cluster studies. Moreover, X-ray properties readily scale with the mass of the host systems \citep[e.g.,][]{kaiser1986}; observations at high-energy are therefore particularly appealing for cosmological studies based on the halo mass function \citep[e.g.][]{borgani1999, henry2009, mantz2010, reiprich2002, ilic2015, bohringer2017, schellenberger2017, pacaud2018}. Definitive assessment of the nature of the emitting sources, including a precise measurement of their redshifts, is provided by the three-dimensional association with their member galaxies, which is achieved in the most robust manner through spectroscopic redshift surveys in the optical domain. These allow measurements at a precision far below the typical velocity dispersions of their host halos.

Collecting large numbers of spectra of candidate member galaxies is an observational challenge requiring highly multiplexed, wide-aperture instrumentation supported by high-quality uniform photometric surveys to draw targets from. The SPIDERS (SPectroscopic IDentification of eROSITA Sources) programme within SDSS-IV \citep{blanton2017} addresses this need in the context of wide-area X-ray surveys. Sharing survey strategies with an observational cosmology project \citep[the extended Baryon Oscillation Survey eBOSS,][]{dawson2016} makes possible the acquisition of a large number of spectroscopic redshifts for objects identified as counterparts of X-ray emitting sources. The main challenge resides in ensuring maximal completeness and wide uniformity of the datasets.

X-ray sources found outside of the Galactic plane can be divided into several categories: galaxy clusters, active galactic nuclei (AGN), X-ray emitting stars, compact objects, etc. \citep[e.g.,][]{voges1999, evans2010, rosen2016}. Each class of object requires its own strategy in order to target the most likely optical counterpart, given the limited positional accuracies, depths, spectral resolution, etc. Numerous AGN candidates found as point-like sources in the R{\"o}ntgen Satellite All-Sky Survey \citep[RASS,][]{trumper1993} and the XMM-Newton Slew survey \citep{saxton2008} require maximum-likelihood or bayesian methods to enhance the chances of association with an infrared or optical AGN \citep{dwelly2017}. Clusters of galaxies are scarcer and they possess extended morphologies on the X-ray sky; however it is difficult to distinguish an extended source from a point source below a certain flux level and beyond the instrumental angular resolution limit. The "red sequence" \citep[e.g.][]{gladders2000}, formed by passive galaxies in those massive halos, helps to characterise them. Multiband photometry of these galaxies provides an estimate of their redshift (e.g.,~by locating the {4000 \AA} break passing through filter passbands). This is the strategy adopted in SPIDERS for selecting clusters: two X-ray samples extracted from the RASS and the XMM-Newton archive are searched for optical red sequences in SDSS DR8 photometry \citep{dr8paper} using the redMaPPer algorithm \citep{rykoff2014}.

A description of the parent SPIDERS galaxy cluster samples and targeting strategies is provided in \citet{clerc2016} (hereafter \citetalias{clerc2016}). The target list was publicly released alongside SDSS Data Release 13 \citep[DR13,][]{dr13paper} in the form of a Value-Added Catalogue\footnote{All VACs are accessible under this link: \url{https://www.sdss.org/dr16/data_access/value-added-catalogs/}.} (VAC). \citetalias{clerc2016} additionally describes the construction of an X-ray selected galaxy cluster sample using spectroscopic data from a pilot survey located in a sub-area of sky covering 300\,$\deg^2$ \citep[see][for a description of the spectroscopic dataset]{dr12paper}. This work demonstrated the end-to-end feasibility of the SPIDERS project; the final catalogue of 230 systems is available as a VAC. This VAC highlighted key features of the sample, including its coverage of the mass-redshift plane and the availability of velocity dispersions. An updated catalogue of 520 systems with high optical richnesses was released in the form of a VAC as part of DR14 \citep{dr14paper} with a corresponding sky area of about $2,500\,\deg^2$. The DR14 catalogue forms the basis of studies on the richness-mass and luminosity-mass relations \citet{capasso2019, capasso2020} and on properties of Brightest Cluster Galaxies \citep{furnell2018, erfanianfar2019}.

The present paper enlarges the scope of the survey and directly relates to DR16 \citep{dr16paper}. This data release contains the entire set of SPIDERS spectra. The motivation of this paper is to provide a detailed census of the data collected in the course of the project, to describe the various steps that led to the final catalogues and to expose salient features in the data, which includes a discussion on the statistical content of the galaxy cluster sample.

The outline of this paper is as follows. Section~\ref{sect:cluster_survey} describes the SPIDERS survey design, the updated targeting strategies, and the sensitivities of the relevant components. Sect.~\ref{sect:cluster_data} provides a detailed assessment of the content of the DR16 spectroscopic data relevant to SPIDERS galaxy clusters. Sect.~\ref{sect:discussion} discusses the adopted survey strategy, and we provide prospects for future massively multiplexed follow-up programmes.
Unless otherwise stated, the cosmological model used in this paper is flat $\Lambda$ Cold Dark Matter with $\Omega_m=0.3$ and $H_0=70$\,km\,s$^{-1}$\,Mpc$^{-1}$. Magnitudes are expressed in their native SDSS (AB) system \citep{fukugita1996}.


\section{The galaxy cluster survey}
\label{sect:cluster_survey}

This section is an overview of the survey design and targeting strategies. We also describe the galaxy cluster selection function prior to spectroscopic observations.

	\subsection{Parent samples and target selection\label{sect:target_selection}}

\begin{figure*}
	\includegraphics[width=\linewidth]{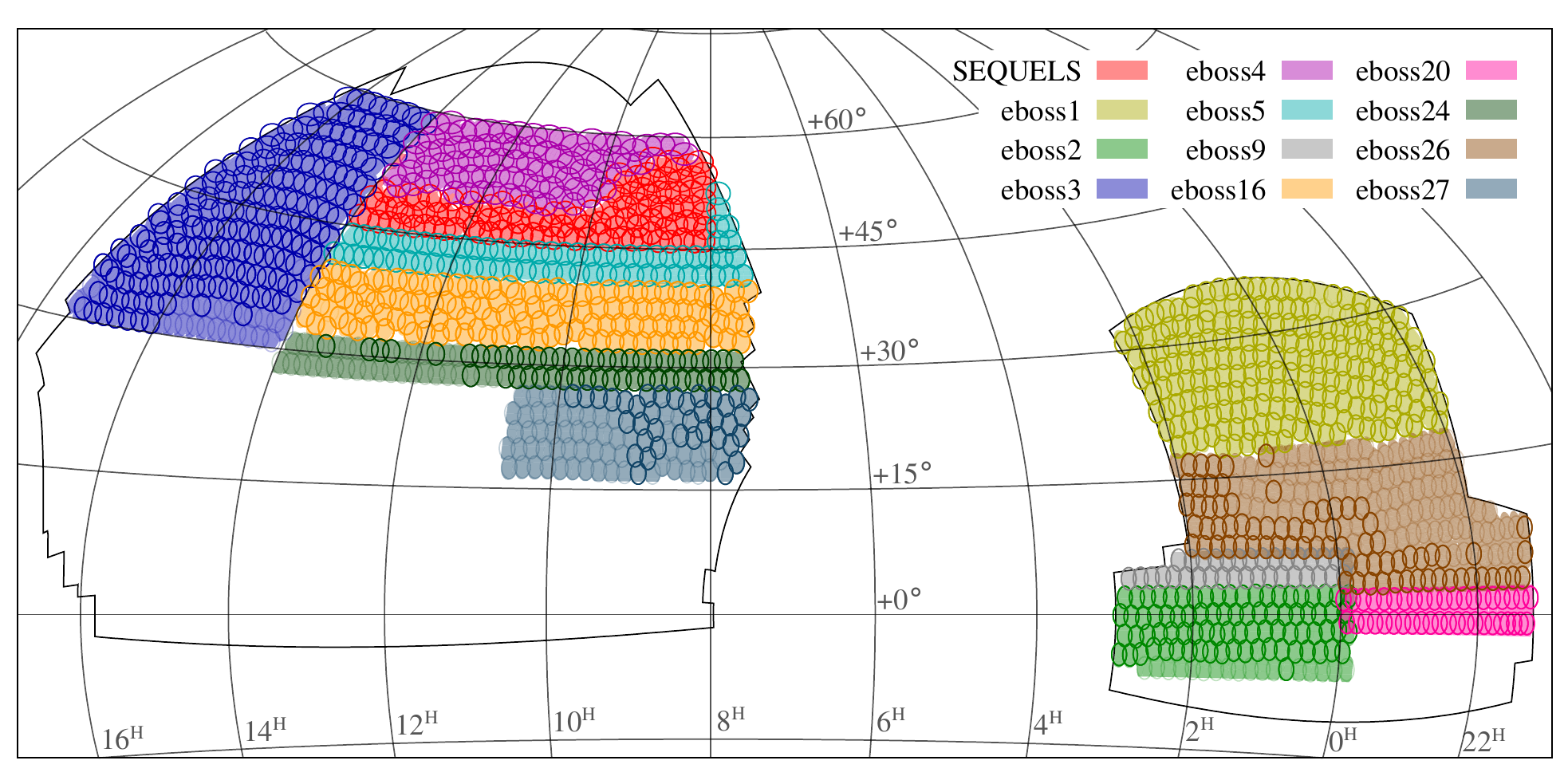}
    \caption{Location of the survey in equatorial coordinates -- this figure updates Fig.~1 in \citetalias{clerc2016}. The thin black contour delineates the BOSS optical imaging area. The various chunks are tiled separately according to target selection algorithms. Chunks with at least one observed plate relevant to the SPIDERS survey are displayed on this figure with colours; they define the SPIDERS targeting area. Circles with a diameter of $3^{\circ}$ represent individual spectroscopic plates. Those outlined with a heavier stroke have average spectral signal-to-noise ratios passing eBOSS requirements (1000 fibres each, shared between eBOSS, TDSS and SPIDERS) and define the SPIDERS DR16 survey area.}
    \label{fig:chunks_sky}
\end{figure*}

We refer the reader to \citetalias{clerc2016} for an in-depth introduction to the CODEX and X-CLASS parent samples and the algorithms applied to select targets in optical imaging data. These targets are astronomical sources identified in photometric catalogues, a vast majority of them are galaxies. Figure~\ref{fig:chunks_sky} displays the area on the sky relevant to this section; i.e.,~the "chunks" \texttt{eboss}1-5, 9, 16, 20, 24, 26, 27 and SEQUELS (\texttt{boss}214, 217) where we effectively selected the targets relevant for the programme and acquired data throughout the survey. All targets described in this section have a target bit mask \texttt{EBOSS\_TARGET1} set to 31 in SDSS products; a few target subclasses were defined, described in the following paragraphs.

		\subsubsection{The CODEX sample and targeting}

CODEX is a search for faint, X-ray extended sources in all-sky \emph{ROSAT} data (RASS), coupling a wavelet algorithm to an automated optical cluster finder in the BOSS $\sim 10\,000\,\deg^2$ $ugriz$ imaging area \citep{finoguenov2020}. Specifically, two different X-ray source catalogues were produced by varying the wavelet threshold and the region of each significant source was searched for an optical counterpart (red-sequence) using the redMaPPer software \citep{rykoff2012, rykoff2014}. The two catalogues were then cross-matched and duplicates were eliminated.

The optical finder provides an estimate for the photometric redshift, $z_{\lambda}$, of the galaxy cluster -- mainly based on the colours of the passive galaxies forming the detected red-sequence -- and an optimized richness estimator $\lambda$, which scales with the number of galaxies exceeding a threshold stellar luminosity. Because of the relatively large uncertainty in the RASS source positions, reaching up to a few arcmin for extended sources, the constraint on the centre is relaxed and the algorithm optimally finds an optical centre within 3\,arcmin of the X-ray position. Refined estimates for the cluster photometric redshift and richness, dubbed 'OPT' (optical), are calculated using the new position.

A critical by-product of this procedure is a list of identified red-sequence members, each assigned a membership probability $p_{\rm mem} \in [0, 1]$. This ranked list of probabilities, attached to each candidate X-ray selected cluster, forms the basis of the targeting strategy and confirmation process \citepalias{clerc2016}.

As explained in \citetalias{clerc2016}, chunk \texttt{eboss3} benefits from a slightly modified targeting scheme due to the higher density of X-ray sources in this area of sky. This scheme effectively favours confirmation of many low-mass systems at the expense of a lower number of members per individual cluster (Sect.~\ref{sect:sample_completeness}).

\begin{figure}
	\includegraphics[width=\columnwidth]{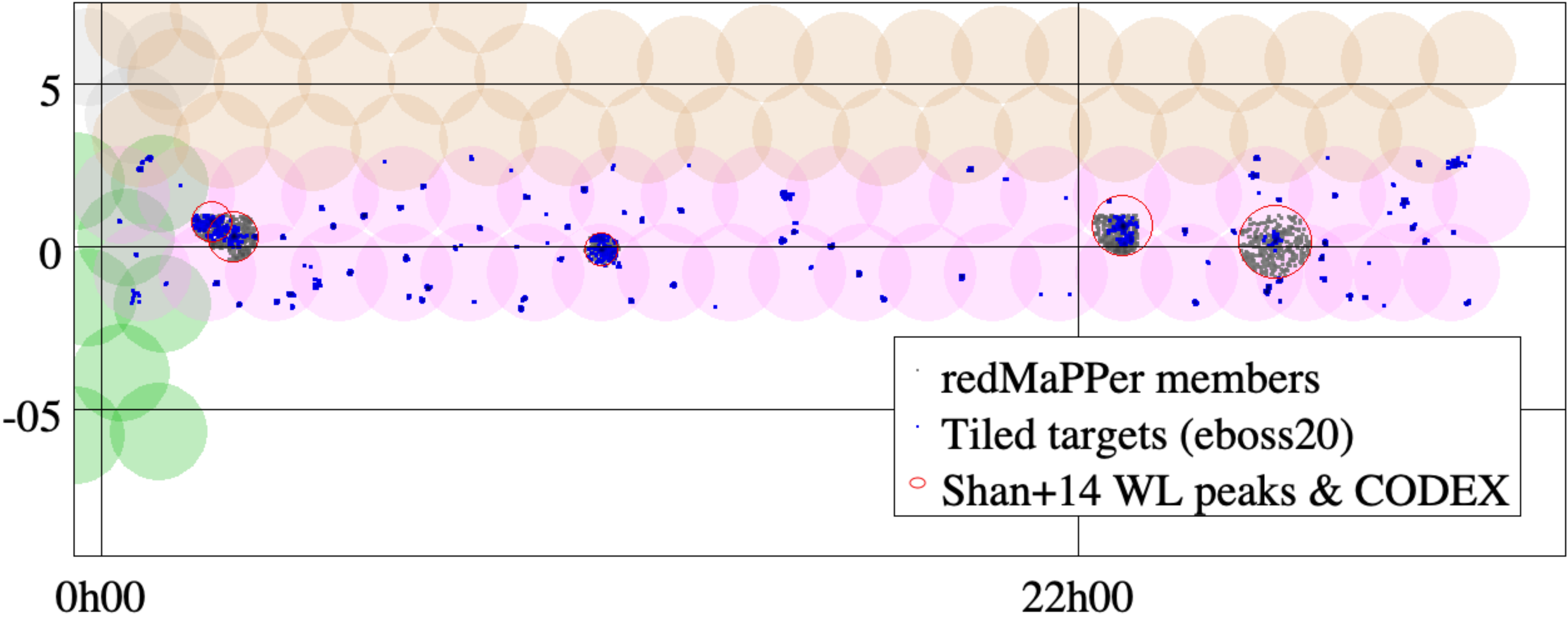}
    \caption{SPIDERS galaxy cluster targets locations in chunk \texttt{eboss20}. This is an expanded view at R.A.$\sim$23h and Dec$\sim 0^{\circ}$ in Fig.~\ref{fig:chunks_sky}. Red circles have radii five times the virial radius of each cluster detected in CODEX and CFHT lensing maps in Stripe82. Photometric members are indicated in light grey; targets indicated as dark blue dots were assigned a spectroscopic fibre.}
    \label{fig:tiling_eboss20}
\end{figure}

In chunk \texttt{eboss20} we added red-sequence targets at larger cluster-centric radii than usual (up to five times the virial radius) in order to enable cluster mass determination through the caustic method \citep{diaferio1999}. A few clusters were selected to this end, specifically weak-lensing detected clusters in the CFHT/Stripe82 imaging survey \citep{shan2014} with a match in the CODEX catalogue. Fig.~\ref{fig:tiling_eboss20} shows the distribution of targets on sky.

All targets described so far have bit mask \texttt{EBOSS\_TARGET2} set to 1.  

A few improvements to the target selection were adopted, mainly intended to help confirm high-redshift systems in chunks \texttt{eboss26} and \texttt{eboss27}. The following additional targets take advantage of deeper optical datasets overlapping the CODEX sample:
\begin{itemize}
 \item \texttt{SPIDERS\_CODEX\_CLUS\_CFHT}: following the procedures described in \citet{brimioulle2013}, pointed CFHT/Megacam observations and CFHT-LS fields provide deep $(u)griz$ photometry. The redMaPPer \citep{rykoff2014} cluster finder detected 598 red-sequence galaxies selected as targets. Among them 515 are new additions to the original list of targets, thereby increasing by about one magnitude the depth probed in the small area of the sky covered by CFHT observations. The corresponding target bit mask \texttt{EBOSS\_TARGET2} is 6.
 \item \texttt{SPIDERS\_CODEX\_CLUS\_PS1}: a few high-redshift ($z_{\lambda}>0.5$) CODEX cluster candidates were searched for red-sequence counterparts in PanStarrs PS1 \citep{flewelling2016}. Our custom version of the Multi-Component Matched Filter (MCMF) tool \citep{klein2018} imposes a radial constraint of 2\,Mpc for members based on the redMaPPer centre. Within this radial cut, each source is assigned a distance in colour-space ($g-r$, $r-i$ and $i-z$) to a red-sequence model. Each source's colour-distance is transformed into a colour-weight in every colour, which is then convolved with a Gaussian smoothing kernel to provide an estimate of the local density of red-sequence galaxies in the region of the cluster candidate. A total of 1,244 members are then selected by applying a threshold to the product of the weight and density within a radius of 1.5\,Mpc from the optical centre with $z < 21$. They contribute 1,194 new objects to the original list of targets. The corresponding target bit mask \texttt{EBOSS\_TARGET2} is 7.
  \item \texttt{SPIDERS\_CODEX\_CLUS\_DECALS}: these targets are output of a custom red-sequence finder code applied to DeCALS photometric data\footnote{\url{http://legacysurvey.org/decamls/}} \citep{dey2019}. 549 photometric members, matching the $g-r$ and $r-z$ colours of early-type galaxies, were selected from the region lying within 0.5\,Mpc of the X-ray position, contributing 517 new targets to the original list. The corresponding target bit mask \texttt{EBOSS\_TARGET2} is 8.
\end{itemize}

		\subsubsection{The X-CLASS sample and targeting}

The targeting of \emph{XMM-Newton} serendipitous sources by SPIDERS is extensively described in \citetalias{clerc2016}. The targeting relies on a posteriori matching of redMaPPer detections down to richness $\lambda=5$ and X-ray "Class 1" sources from the X-CLASS survey \citep{sadibekova2014}. Target priorities were boosted relative to CODEX clusters to enhance their chance of being observed. The target bit mask \texttt{EBOSS\_TARGET2} is set to 5.
Differences with respect to the CODEX sample are the scarcity of X-CLASS sources (pointed \emph{XMM} observations cover a small sky area), the excellent positional accuracy of the X-ray detections (below $10\arcsec$), and the uncontaminated nature of the sample (composed of truly extended X-ray objects). These characteristics make the spectroscopic confirmation of X-CLASS sample more straightforward.

	\subsection{Survey design\label{sect:survey_design}}

\begin{figure*}
	\includegraphics[width=\linewidth]{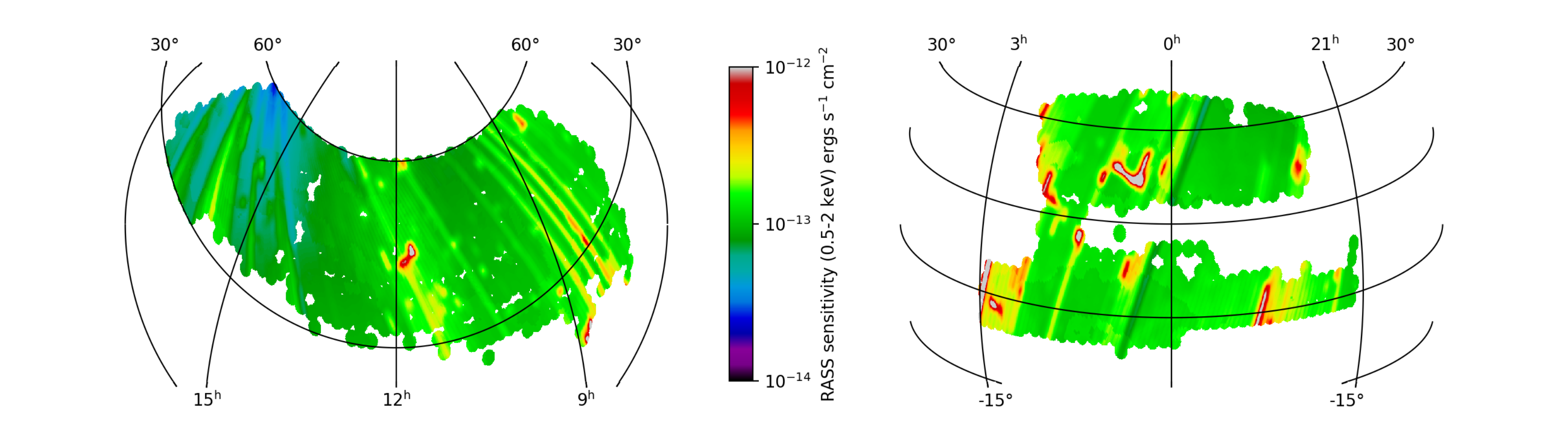}
    \caption{The X-ray sensitivity for CODEX detections in the SPIDERS footprint (5,350\,deg$^2$) shown in equatorial coordinates, in the northern (left) and southern (right) Galactic caps. Equal-area pixel binning on the sphere is performed on a $7\arcmin$ scale ($N_{\rm side}=512$). The colour scale is logarithmic; gray areas indicate regions of very low sensitivity or unusable X-ray data. Individual eBOSS/SPIDERS spectroscopic plates are recognisable from their circular, $3^{\circ}$ diameter footprint.}
    \label{fig:sky_coverage}
\end{figure*}

Figure~\ref{fig:sky_coverage} presents the sky coverage achieved after five years of survey, including the SEQUELS area acquired as part of SDSS-III \citep{dr13paper}. The colour coding indicates the X-ray sensitivity of the ROSAT all-sky survey (RASS) from which the CODEX detections were drawn. The pixel size is approximately $7\arcmin$, equivalent to $N_{\rm side}=512$ in the HEALPix\footnote{\url{http://healpix.sourceforge.net}} sky pixelation scheme. Despite local inhomogeneities, the survey depth is rather uniform around $1-2 \times 10^{-13}$\,ergs\,s$^{-1}$\,cm$^{-2}$ in the (0.5 -- 2)~keV band. Note, however, the slight increase in depth at R.A.\,$\sim 16$\,h, $\delta \sim +50 ^{\circ}$, close to the deep ROSAT North Ecliptic Pole (itself at R.A.\,$=18$h, $\delta=+66^{\circ}$). As mentioned in the previous section, a modified targeting scheme is used to take advantage of the higher source number density in this area of sky.

In Fig.~\ref{fig:sky_coverage}, each circle actually represents the footprint of an eBOSS spectroscopic plate. Only those plates acquired with average spectral signal-to-noise ratios reaching the quality threshold imposed by the survey are considered. The survey geometrical area subtends 5,350\,deg$^2$. Each of these 1,137 plates has diameter $3^{\circ}$ (7\,$\deg^2$ geometrical area). Neighbouring plates substantially overlap; almost half of the geometrical area covered by one plate is common to one or several other plates. The plate distribution is slightly denser in the SEQUELS area~\citepalias{clerc2016}.

Each spectroscopic plate accommodates 1,000 spectroscopic fibres linked to the BOSS spectrographs \citep{smee2013} on the SDSS telescope \citep{gunn2006}; a small fraction (a few percent) is dedicated to SPIDERS galaxy cluster targets, because of the low sky density of CODEX candidates (less than one per square degree). Most of the fibres are dedicated to the eBOSS programs (Quasars, Luminous Red Galaxies, Lyman-$\alpha$ emitters, etc.), with some assigned to SPIDERS AGN targets and Time-Domain Spectroscopic Survey \citep{macleod2018} sources. Because of the rarity of X-ray sources and the constraint imposed by strongly clustered SPIDERS galaxies, these objects received high priority in the process of assigning fibres to targets. Such a strategy ensured that a maximum of cluster targets could be assigned a fibre, especially at places where plates overlap, despite the large fibre collision radius ($62\arcsec$) relative to arcminute-sized galaxy clusters.

	\subsection{Survey sensitivities prior to spectroscopic follow-up\label{sect:survey_sensitivities}}

		\subsubsection{The CODEX survey depths}

\begin{figure}
	\includegraphics[width=\columnwidth]{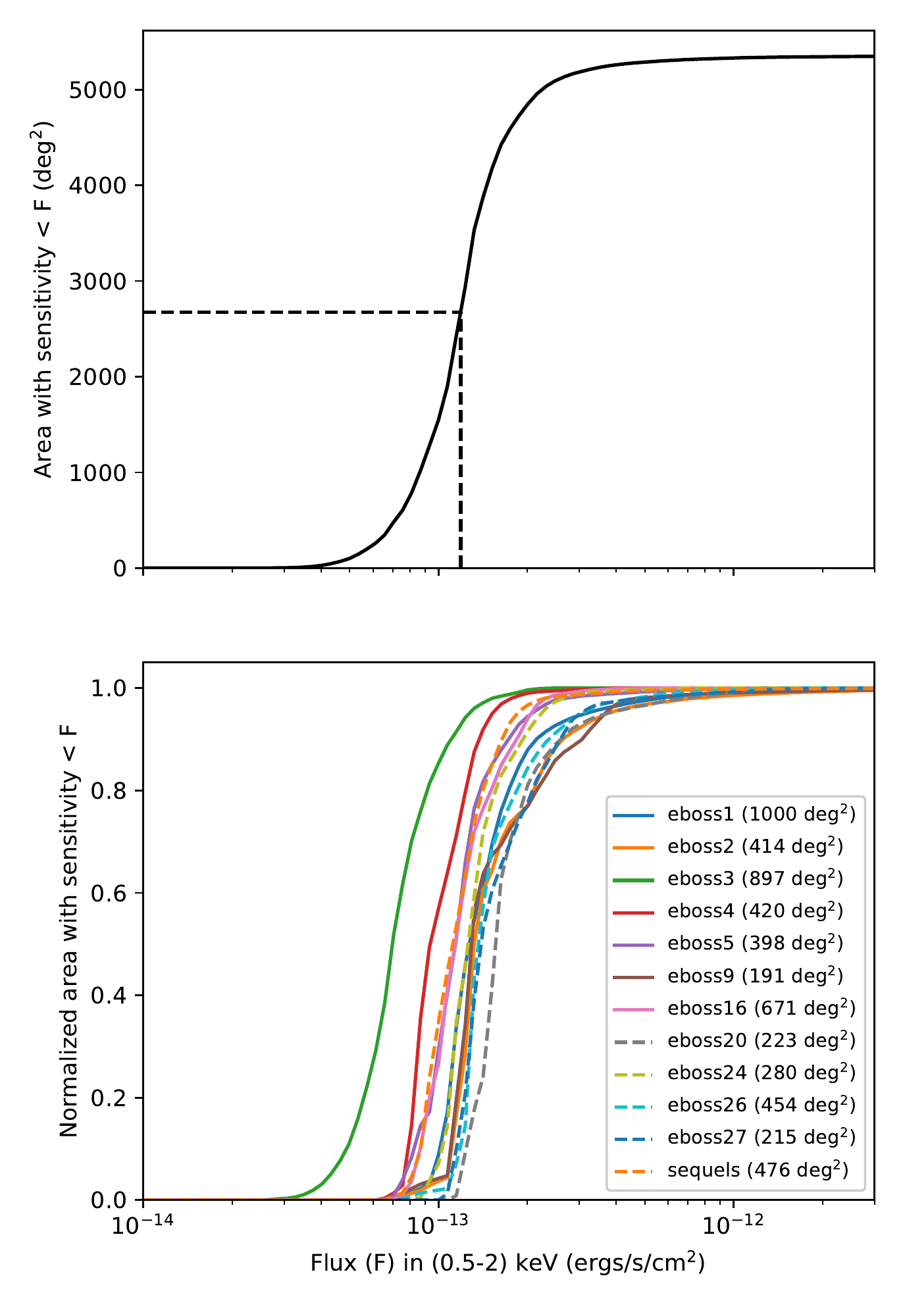}
    \caption{X-ray flux sensitivity curves for the CODEX survey within the SPIDERS footprint, based on the $N_{\rm side}=512$ map in Fig.~\ref{fig:sky_coverage}. The top panel shows the sky area area covered as a function of limiting flux. The dashed lines indicates the median $\sim 10^{-13}$\,erg\,s$^{-1}$\,cm$^{-2}$ sensitivity (in the band 0.5--2\,keV). The bottom panel is a breakdown into the various sky chunks (displayed in Fig.~\ref{fig:chunks_sky}). All curves are normalised by the area covered by the spectroscopic survey, indicated in the legend.}
    \label{fig:areacurve_codex}
\end{figure}

The area sensitivity curves for the CODEX X-ray detections in the SPIDERS footprint is displayed in Fig.~\ref{fig:areacurve_codex}. The bottom panel splits the visualization across the various chunks (shown in Fig.~\ref{fig:chunks_sky}). Such partitioning is arbitrary since the shapes of chunks were designed independently from any X-ray sensitivity consideration. This figure highlights the relatively narrow spread of their median flux limit, $f_{{\rm lim}, 50\%}$, around the value $1.2 \times 10^{-13}$\,erg\,s$^{-1}$\,cm$^{-2}$ among the multiple chunks, except for the \texttt{eboss3} area which is slightly deeper ($f_{{\rm lim}, 50\%} = 0.7 \times 10^{-13}$\,erg\,s$^{-1}$\,cm$^{-2}$).

\begin{figure*}
	\includegraphics[width=\linewidth]{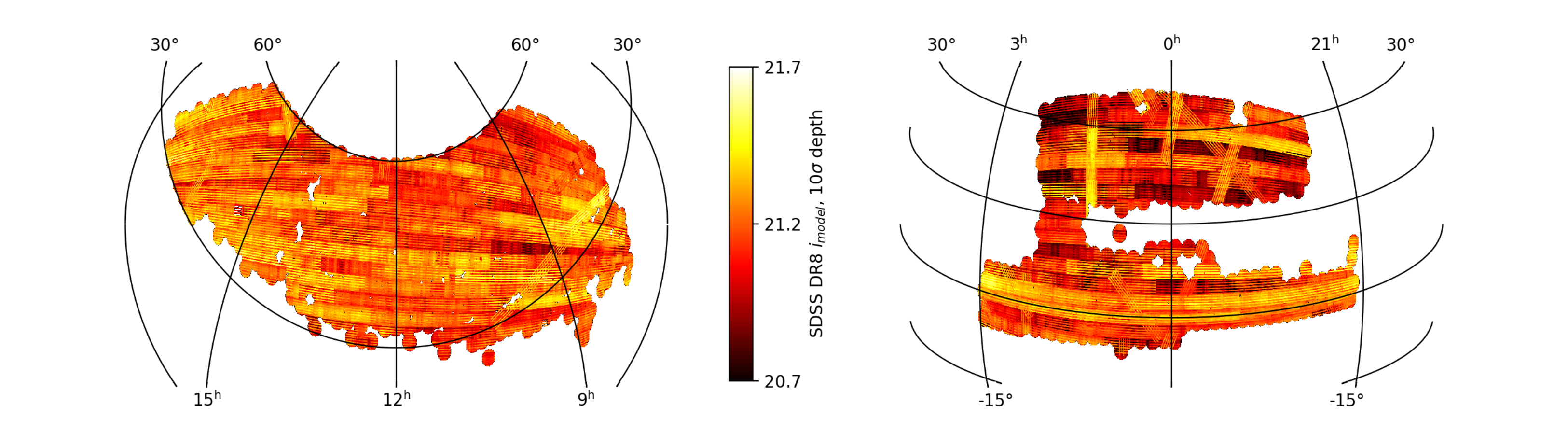}
    \caption{The SDSS-$i$ band photometric sensitivity for redMaPPer detections in the SPIDERS footprint (5,350\,deg$^2$). This figure is adapted from \citet{rykoff2015}. The colour scale spans the $\pm 3$ standard-deviation range ($\sigma_i = 0.16$\,mag on the $\sim 1.7 \arcmin$ pixel scale of these maps) around the median value $i_{\rm model}=21.2$. Projection and coordinates are identical to those in Fig.~\ref{fig:sky_coverage}.}
    \label{fig:sky_coverage_imodel}
\end{figure*}

\begin{figure}
	\includegraphics[width=\columnwidth]{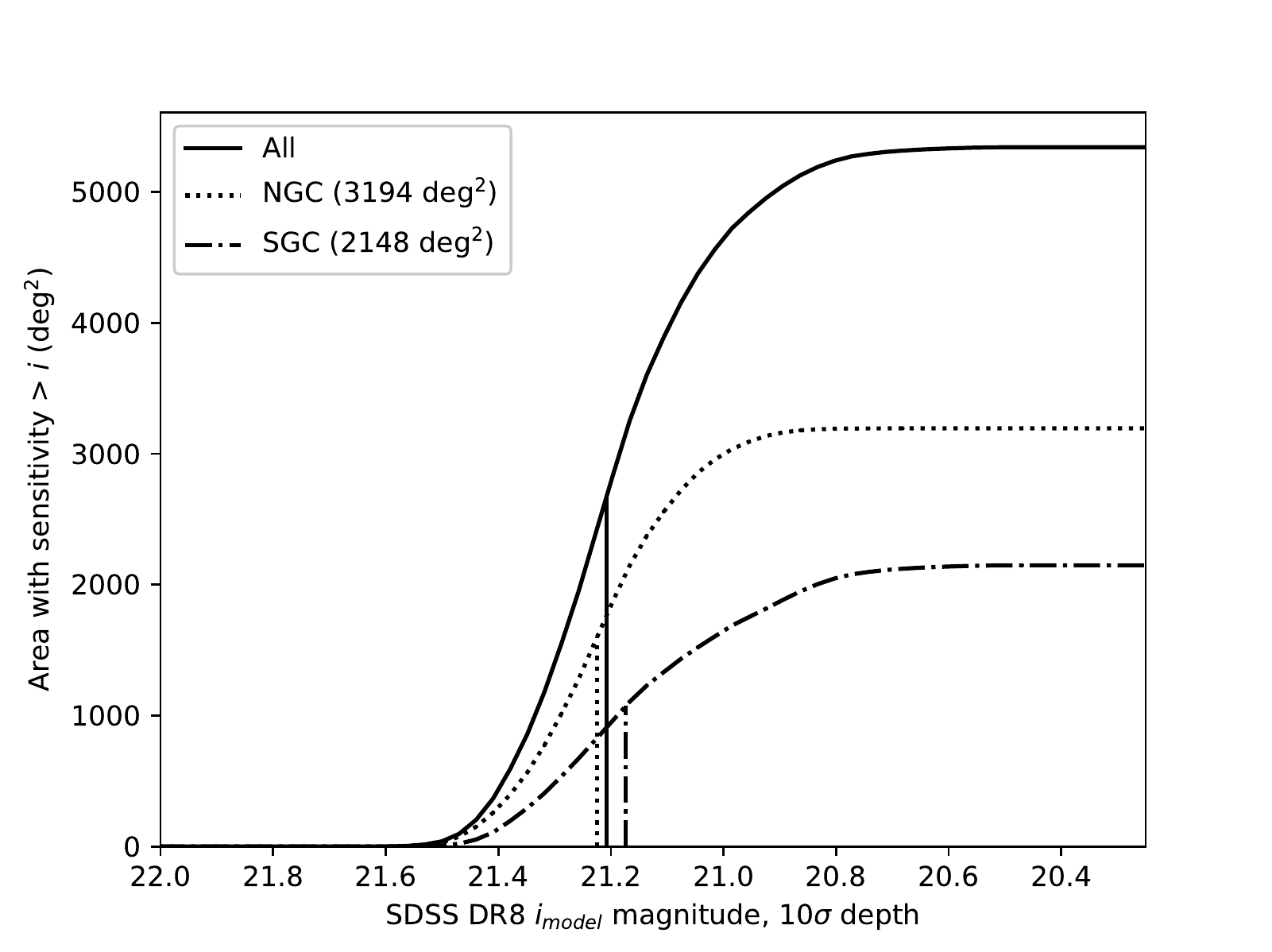}
    \caption{Photometric sensitivity area curves in the SDSS $i_{\rm model}$ band used for redMaPPer detections in the SPIDERS survey. The southern Galactic cap (SGC) is slightly shallower than the northern Galactic cap (NGC) on average, with a larger fraction of shallower sky patches. $i$-band sensitivity maps were filtered on a $\sim 1.7\arcmin$ scale to produce these curves.}
    \label{fig:areacurve_redmapper}
\end{figure}

Figure~\ref{fig:sky_coverage_imodel} shows the sensitivity (the 10$\sigma$ depth in the $i_{\rm model}$ band for galaxy sources\footnote{"Model" magnitudes are optimal measures of the flux using a matched galaxy model \citep{stoughton2002}.}) of the SDSS DR8 photometry on a scale of $\sim 1.7 \arcmin$ ($N_{\rm side}=2048$) restricted to the SPIDERS footprint. This map was extracted from the redMaPPer maps in HEALPix format\footnote{Retrievable at \url{http://risa.stanford.edu/redmapper/}} and is relevant for the version of the data and algorithms used in this work. The procedure followed in constructing these maps, as well as a discussion on their peculiar features, is given in \citet{rykoff2015}. It is beyond the scope of this paper to provide details of these photometric maps; rather, we emphasize here the typical $\sim$0.2\,mag depth variation across the survey in the $i$ band. For the purpose of identifying counterparts to faint ROSAT sources, this is sufficiently uniform coverage. There is, however, a noticeable difference between the North Galactic Cap (NGC) and South Galactic Cap (SGC) parts of the sky, which impacts the spectroscopic observations of higher-redshift and lower-mass systems. This is more easily seen in Fig.~\ref{fig:areacurve_redmapper}, where the pixelated maps of Fig.~\ref{fig:sky_coverage_imodel} were used to compute the effective area as a function of limiting magnitude. The SGC has a slightly higher median limiting magnitude, and a tail towards shallower sky patches.

\begin{figure*}
	\includegraphics[width=\linewidth]{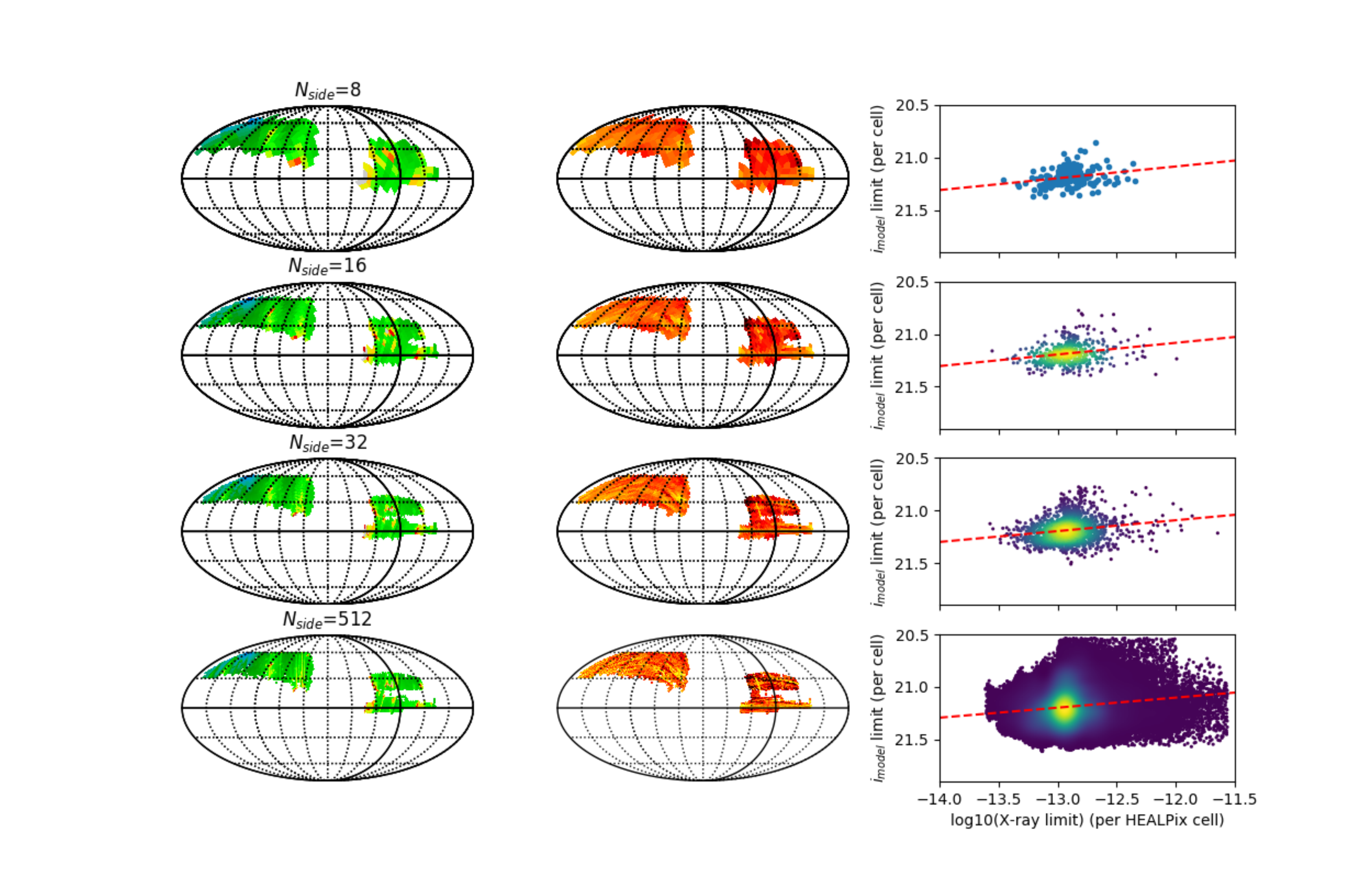}
    \caption{A mild correlation between sensitivities in the optical and X-ray bands is observed in the survey footprint. Maps in the first and second column are essentially similar to Fig.~\ref{fig:sky_coverage} and Fig.~\ref{fig:sky_coverage_imodel}, respectively (identical color scales and coordinate systems, but Mollweide projection). They are displayed at various increasing resolutions (from top to bottom) indicated by the value $N_{\rm side}$, using average pixel binning (see text). The rightmost panel shows the pixel-to-pixel correspondance between the X-ray limiting flux and the $i$-band magnitude limit, colour coded by density. A linear model is fit to the datapoints as an indication of the general trend. It provides similar results at all tested resolutions, namely $i_{\rm model, lim} = 21.2 -0.1  \log_{10}(f^X_{\rm lim}/1.2\times10^{-13} {\rm cgs})$.}
    \label{fig:corr_xsens_osens}
\end{figure*}

The sensitivities (both X-ray and optical) are essentially uniform across the survey area, which eases the calculation of a selection function for cosmology.
Figure~\ref{fig:corr_xsens_osens} explores spatial correlations between the X-ray and $i$-band sensitivities across the survey area. We performed pixel-alignment and downsampled the HEALPix maps (Fig.~\ref{fig:sky_coverage} and~\ref{fig:sky_coverage_imodel}) at various resolutions ($N_{\rm side} = 2^3, 2^4, ..., 2^9$) by averaging sensitivities in adjacent pixels and maintaining the equal-area HEALPix gridding. Such a binning scheme is a crude approach in the presence of strong gradients in the sensitivity maps (e.g. holes in the X-ray exposure map); however, those features are relatively few. Matching the resulting maps pixel-to-pixel reveals that the X-ray and optical sensitivities are weakly correlated at all studied resolutions (third panel in the figure). Fits to the data points provide similar results at resolutions corresponding to all values of $N_{\rm side}$, with a slope $(-0.1)$ and intercept 21.2 when expressed in terms of $i_{\rm model, lim}$ and $\log_{10}(f^X_{\rm lim}/1.2 \times 10^{-13} {\rm cgs})$. 
The absolute Spearman correlation coefficient ranges between 0.24, in the low-resolution case, and 0.12, in the high-resolution case; the p-value remains very small due to the high number of points. This analysis portrays a weak, albeit significant, correlation between the optical and X-ray survey sensitivities.  This feature has been exploited in the targeting strategy and the spectroscopic confirmation of cluster candidates, for instance by imposing a uniform target magnitude limit ($i_{\rm fiber2}=21.2$) and performing visual validation of candidates independently from the exact location of a candidate on sky.

		\subsubsection{The X-CLASS survey depths}

\citet{clerc2012} presented a method for deriving the selection function of sources selected as 'C1' extended sources in the XMM-Newton archive. This technique is supported by extensive end-to-end simulations of individual pointings coupled with the same source detection pipeline as in the data analysis. The `C1' class \citep{pacaud2006} is defined as an almost pure selection of extended sources based solely on X-ray detection criteria (measured extent, extent likelihood specifically). In this approach, `pure' means free from spurious detections and contamination by point-like sources, as long as the input conditions of the simulations are a faithful representation of real observing conditions. Overlaps of neighbouring pointings (each spanning $\sim 30\arcmin$ on sky) are treated consistently with the catalogue construction (e.g.~higher exposure pointings are favoured, while two pointings of similar exposure bear equal weights). In practice, selection functions pre-computed on a grid of pointing exposure time, local background level, and local Galactic absorption are interpolated and this provides the probability of detecting a galaxy cluster of core-radius $R_c$ (arcsec) and flux $f_X$ (in the 0.5--2 keV energy band) at a given sky location.
The X-CLASS survey is significantly deeper than RASS; for instance, a galaxy cluster represented as a $\beta$-model \citep{cavaliere1976} surface brightness profile with apparent core-radius $20\arcsec$ and (0.5--2)\,keV flux $3\times 10^{-14}$\,erg\,s$^{-1}$\,cm$^{-2}$ has about $50\%$ probability of being detected and classified as `C1' in a 10\,ks XMM observation with standard background levels.

Because the same imaging dataset is used in finding counterparts to X-CLASS sources as to CODEX sources, optical sensitivities are identical to those presented previously (e.g.~Fig.~\ref{fig:sky_coverage_imodel}); in particular, these optical sensitivities limit the accessible range of cluster redshifts to $0 \lesssim z \lesssim 0.7$. The comparatively deeper X-ray survey thus enables probing lower mass clusters on average, almost reaching the group regime. Finally, recall that the matching procedure between X-ray and optical cluster detections slightly differs from that used to build the CODEX sample: it relies on a cross-match between two independent catalogues, while CODEX is a red-sequence search around each X-ray source. It is beyond the scope of this paper to quantify the subsequent selection introduced by the optical/X-ray cross-matching; the reader is referred to \citet{sadibekova2014} for further details.


\section{Galaxy cluster survey data}\label{sect:cluster_data}

This section provides an overview of the final spectroscopic data and products obtained through the SPIDERS survey. We also review the main features of the galaxy cluster catalogues constructed from this sample, described in dedicated papers: \citet{kirkpatrick2020} (hereafter \citetalias{kirkpatrick2020}) for CODEX and \citet{xclass2020} for X-CLASS.

	\subsection{Tiling and fiber assignment\label{sect:tiling}}

Tiling is the process that assigns spectroscopic fibres to targets in the most efficient manner. Survey tiling is performed simultaneously for all eBOSS subcomponents, including SPIDERS \citep{dawson2016}; each chunk being treated independently from the others. Each survey subcomponent provides a list of prioritised targets to the algorithm, such targets are denoted as `submitted'. A submitted target may be present in more than one list. Targets which are assigned a spectroscopic fibre are denoted as `tiled'; these targets enter an observing queue at the telescope and they may eventually be observed. \citetalias{clerc2016} provides a presentation of the tiling results with a perspective on galaxy cluster confirmation efficiency and the impact of the fibre collision radius. Most of these findings are unchanged for the chunks most recently tiled (\texttt{eboss20/24/26/27}). Due to both the requirement of high completeness and the low number density of sources, SPIDERS targets are assigned fibres with highest priorities; putative Brightest Central Galaxies are prioritised first among those. Besides the natural dimming of cluster galaxies with increasing redshift, strong observational constraints are imposed by the $62\arcsec$ collision radius of fibres feeding the BOSS spectrographs. Typical SPIDERS galaxy cluster extents range from two to a few arcmin, and red-sequence galaxies are concentrated towards the cluster centre as their luminosity increases. Overlaps of spectroscopic plates allows one to revisit a certain area of sky twice (or more), thus slightly alleviating the issue of fibre collision.

\begin{figure}
	\includegraphics[width=\columnwidth]{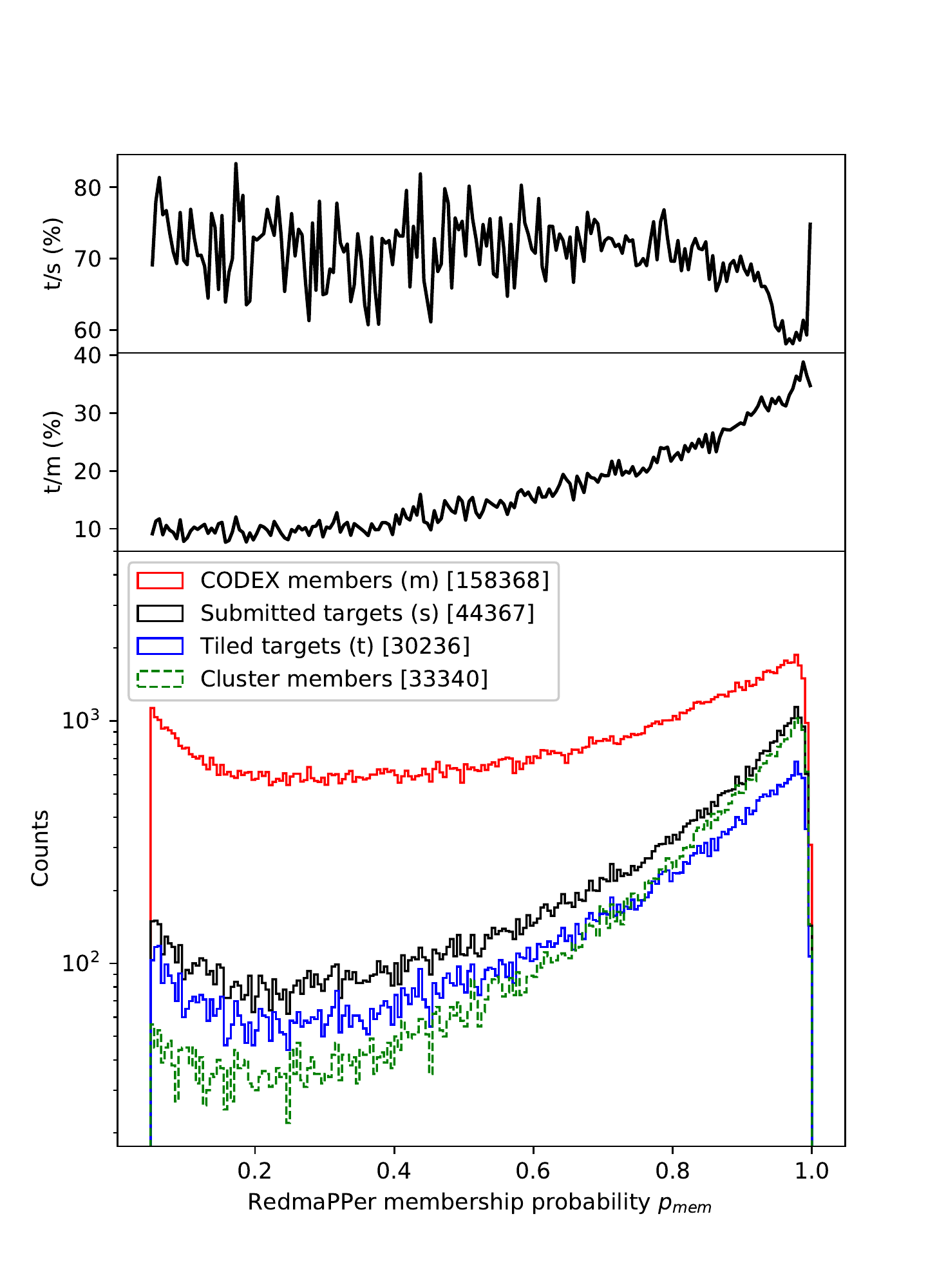}
    \caption{Histograms of photometric cluster membership probabilities $p_{\rm mem}$ for SPIDERS targets. The top panel represents the target allocation completeness (ratio between tiled and submitted targets). The middle panel represents the target selection completeness (ratio between tiled and all targets). The bottom panel contains histograms for SPIDERS-submitted targets (black curve) and tiled targets in SPIDERS clusters (blue curve) computed in bins of width $\Delta p = 0.005$. The top curve (in red) represents all CODEX cluster members identified by redMaPPer with a $p_{\rm mem} > 0.05$. The green dashed curve represents galaxies spectroscopically identified as members of the X-ray cluster catalogue (\citetalias{kirkpatrick2020}, see also Sect.~\ref{sect:spicatalogue}). It includes spectra acquired prior to SPIDERS; hence their number exceeds the number of tiled targets. Numbers in brackets are the totals for each histogram.}
    \label{fig:tiling_probabilities}
\end{figure}

Figure~\ref{fig:tiling_probabilities} presents numbers of targets grouped by their redMaPPer membership probability ($p_{\rm mem}$). The histograms are computed for the area of sky surveyed by SPIDERS (5,350\,deg$^2$ as shown in Fig.~\ref{fig:sky_coverage}). Roughly speaking, $p_{\rm mem}$ encapsulates information about the luminosity of a galaxy, its distance in colour-space from the red-sequence, and its physical distance to a cluster centre. By design of the targeting algorithm, $p_{\rm mem}$ is always greater than 0.05 for SPIDERS clusters.
The target allocation completeness (shown in the top panel) is roughly constant at $\sim 70\%$ up to $p_{\rm mem} \sim 0.8$. The allocation completeness slightly decreases at high $p_{\rm mem}$ as a consequence of the finite fibre collision radius and the spatially concentrated nature of high-$p_{\rm mem}$ galaxies within a cluster. The middle panel represents the target selection completeness (ratio between the blue and red curves). The target selection algorithm is able to allocate a fibre to 10\% of all photometric members with $p_{\rm mem} \lesssim 0.4$. This ratio continuously rises up to $\sim 35\%$ as $p_{\rm mem}$ approaches 1.

The number of tiled targets displayed in Fig.~\ref{fig:tiling_probabilities} (30,236) is slightly higher than the actual number of observed targets (29,609, see next section), even though they share identical sky footprints. Unobserved targets are of two kinds. Those located at the border of the survey footprint were assigned a fibre on a neighbouring spectroscopic plate that was not observed during the course of the survey (see also Fig.~\ref{fig:2d_completeness}). Those located in the SEQUELS footprint were initially assigned a fibre and were discarded while applying the SPIDERS target algorithm \citepalias{clerc2016}.
We also show in Fig.~\ref{fig:tiling_probabilities} the number of spectroscopically identified cluster members (see Sect.~\ref{sect:spicatalogue}). Their total number exceeds the number of tiled targets, because our final analysis makes use of data obtained prior to SPIDERS during previous SDSS projects.

	\subsection{Spectral quality and completeness of SPIDERS spectra\label{sect:spectral_quality}}

In this section we discuss specifically new targets acquired as part of the SPIDERS galaxy cluster survey. 
From the SDSS DR16 spectroscopic sample, we retrieve all those targets with bit flag \texttt{EBOSS\_TARGET1} set to 31 (SPIDERS targets) and \texttt{EBOSS\_TARGET2} set to 1 (CODEX cluster targets) or 5 (X-CLASS cluster targets). The overlap between the CODEX and X-CLASS cluster samples is non-zero, hence the overlap between the corresponding target classes.
We also include in this discussion galaxy cluster targets acquired as part of the SEQUELS survey, released together with SDSS DR13 \citep{dr13paper} and reanalysed in DR16; these are obtained by selecting objects with bit flags \texttt{EBOSS\_TARGET0} and \texttt{ANCILLARY\_TARGET2} set to 21 and 53, respectively. To ensure removal of duplicate acquisitions, only objects with \texttt{SPECPRIMARY} set to 1 are considered.

Table~\ref{tab:spectral_bdown} summarises the numbers of spectra obtained as part of the SPIDERS galaxy cluster program. The top section of this table relates to SPIDERS targets, the bottom to SEQUELS targets. The vast majority of SPIDERS cluster spectra are linked to a CODEX (RASS) target. A small overlap ($0.4\%$) with the SPIDERS AGN targets is noticeable, due to the different nature of the algorithms matching X-ray sources to optical sources for clusters and AGN.
Consistent with the high overall signal-to-noise ratio of the spectra, only $\sim 1 \%$ of the spectra have a \texttt{ZWARNING\_NOQSO} > 0, indicative of a less reliable fit. Those spectra correspond to low signal-to-noise ratio spectra or absence of signal in the detectors, due e.g., to unplugged fibres.

\begin{table}
	\centering
	\caption{Breakdown of the SPIDERS galaxy cluster targets in the SDSS DR16 spectroscopic dataset. Target classes are not mutually exclusive; for instance, a number of SPIDERS AGN targets \citep{dwelly2017} are also SPIDERS cluster targets. The bottom section refers to SEQUELS galaxy cluster spectra partially described in \citetalias{clerc2016}, released in SDSS DR13 and reprocessed as part of DR16. The overlap between SPIDERS and SEQUELS spectra is null.
	Notes: $^{(a)}$: defined by \texttt{ZWARNING\_NOQSO} = 0. $^{(b)}$: corresponding to the best-fit "NOQSO" template.}
	\label{tab:spectral_bdown}
	\begin{tabular}{lrr}
		\hline
		SPIDERS clusters spectra & 26817 & \\
		- CODEX targets & 26016 & 97.0\% \\
		- XCLASS targets & 1134 & 4.2\% \\
		- SPIDERS AGN targets & 121	& 0.4\% \\
		- Reliable fit$^{(a)}$ & 26508 & 98.8\%\\
		- Reliable fit$^{(a)}$, type$^{(b)}$ "GALAXY" & 26399 & 98.4\% \\
		\ * with $0 < z < 0.3$ &	16940 & (64.2\%) \\
		\ * with $0.3 < z < 0.6$ & 9382	& (35.5\%) \\
		\ * with $z > 0.6$	& 77	& (0.3\%) \\
		\hline
		\hline
		SEQUELS clusters spectra & 2792	&	\\
		- Reliable fit $^{(a)}$ &  2769 & 99.2\%	\\
		- Reliable fit$^{(a)}$, type$^{(b)}$ "GALAXY" & 2761 & 98.9\% \\
		\ * with $0 < z < 0.3$ & 1850 & (67.0\%) \\
		\ * with $0.3 < z < 0.6$ & 909	& (32.9\%) \\
		\ * with $z > 0.6$	& 2	& ($<0.1$\%) \\		\hline
	\end{tabular}
\end{table}

\begin{figure}
	\includegraphics[width=\columnwidth]{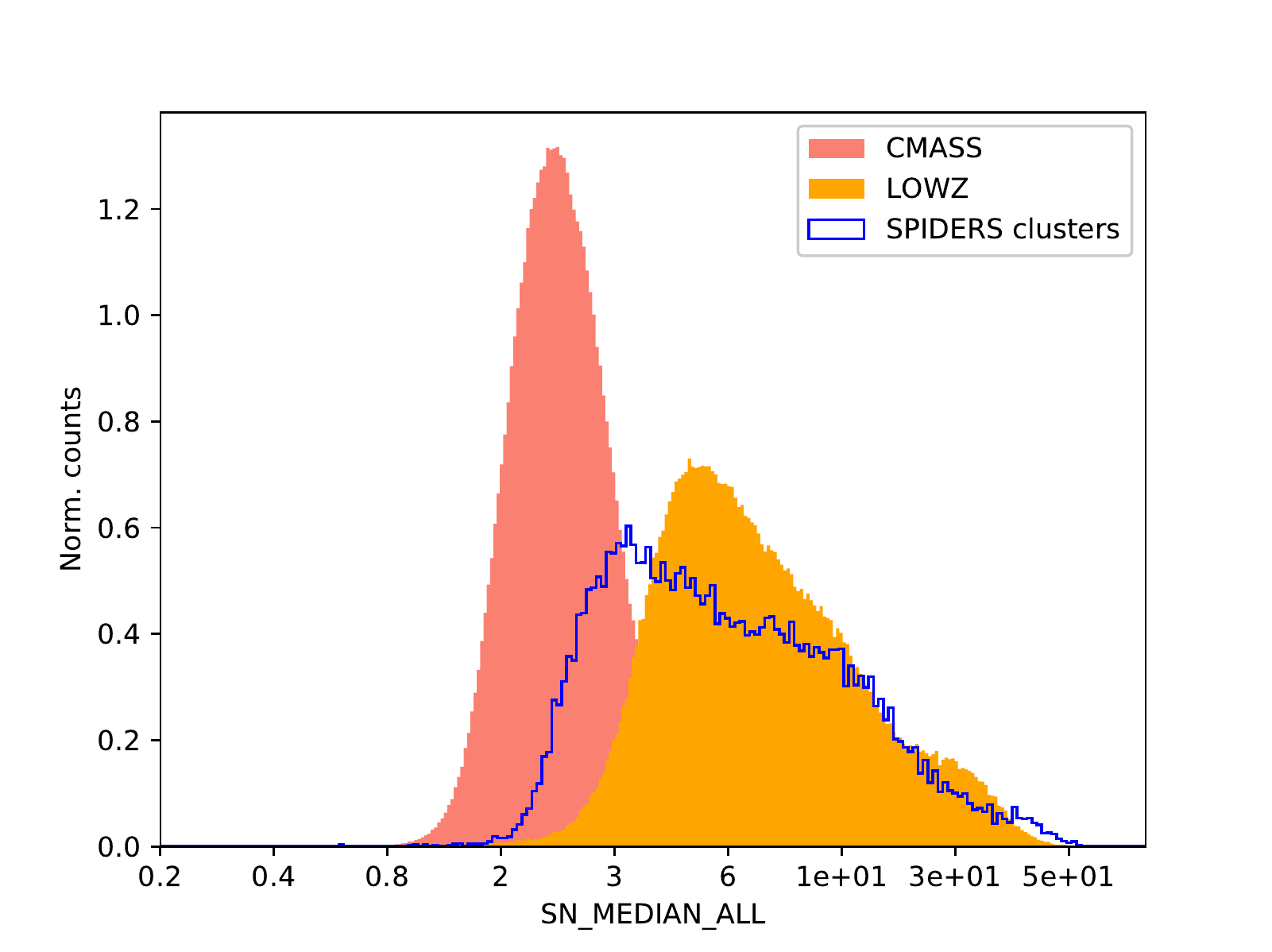}
    \caption{Distribution of the median spectral signal-to-noise ratio (\texttt{SN\_MEDIAN\_ALL}) for all SPIDERS targets. As a comparison, the distributions for the BOSS CMASS and BOSS LOWZ samples are also shown.}
    \label{fig:distrib_sn_median}
\end{figure}

\begin{figure}
	\includegraphics[width=\columnwidth]{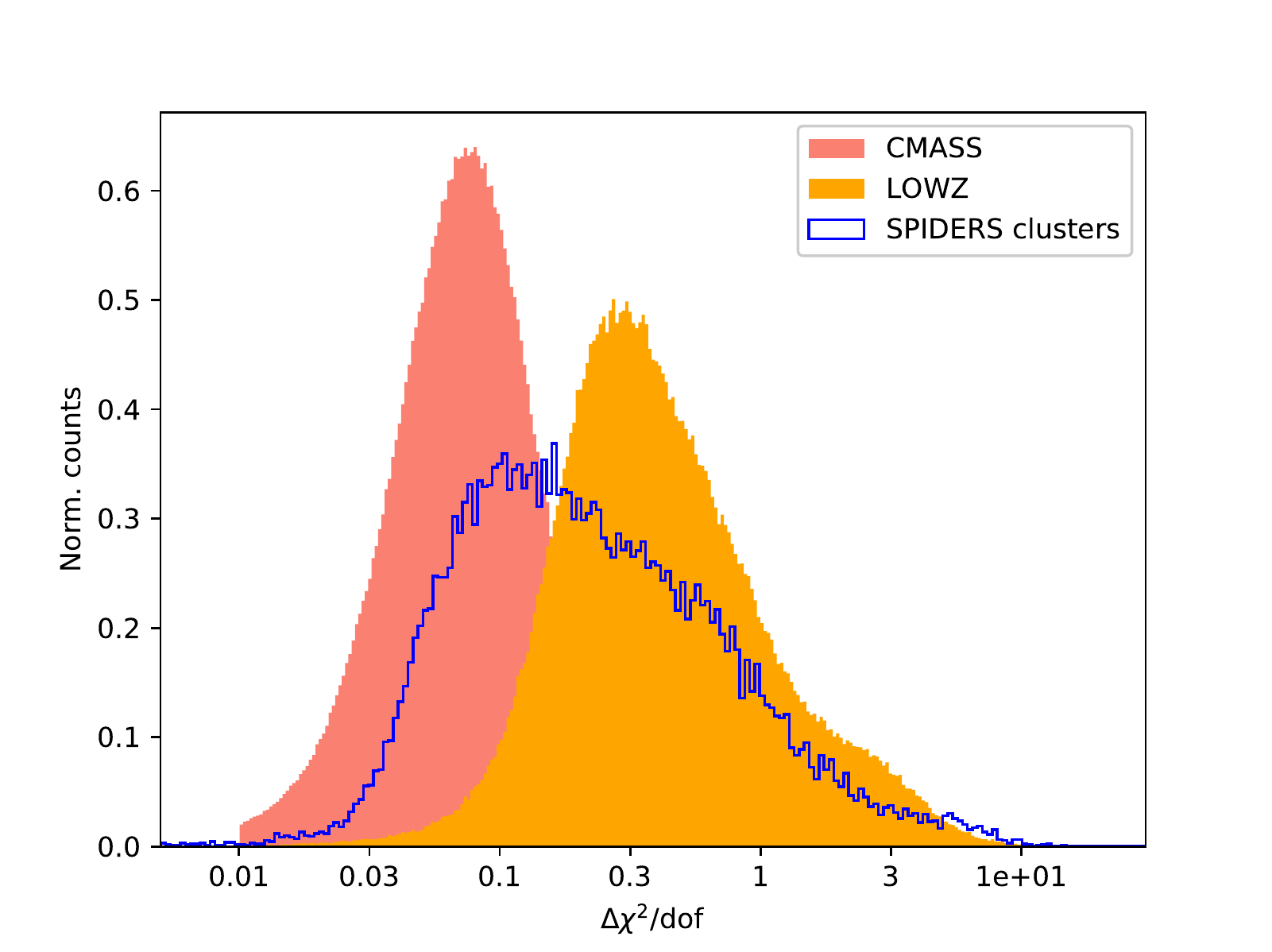}
    \caption{Distribution of $\Delta \chi^2/{\rm dof}$ for all SPIDERS targets. This quantity represents the difference between the best fit template and second best fit template after exclusion of quasar templates from the fit. As a comparison, the distributions for the BOSS CMASS and BOSS LOWZ samples are also displayed.}
    \label{fig:distrib_rchi2diff_noqso}
\end{figure}

The histograms in Fig.~\ref{fig:distrib_sn_median} and Fig.~\ref{fig:distrib_rchi2diff_noqso} provide more detailed views of the quality of this new spectroscopic dataset in terms of the median signal-to-noise ratio of the spectra (\texttt{SN\_MEDIAN\_ALL}) and the $\Delta \chi^2/{\rm dof}$ value, respectively. The former is interpreted as a tracer of the quality of the spectra; the latter is the difference between the reduced $\chi^2$ of the best fit template and the second best fit template after exclusion of quasar templates from the fit and is therefore an indicator of the reliability of the template selected to determine the redshift of the target \citep{bolton2012}. To ease the comparison, the distributions for the widely-used CMASS and LOWZ galaxy samples of SDSS-III/BOSS \citep{dawson2013} are displayed; SPIDERS spectral quality is located between these two samples and its distribution is broader.

\begin{figure}
	\includegraphics[width=\columnwidth]{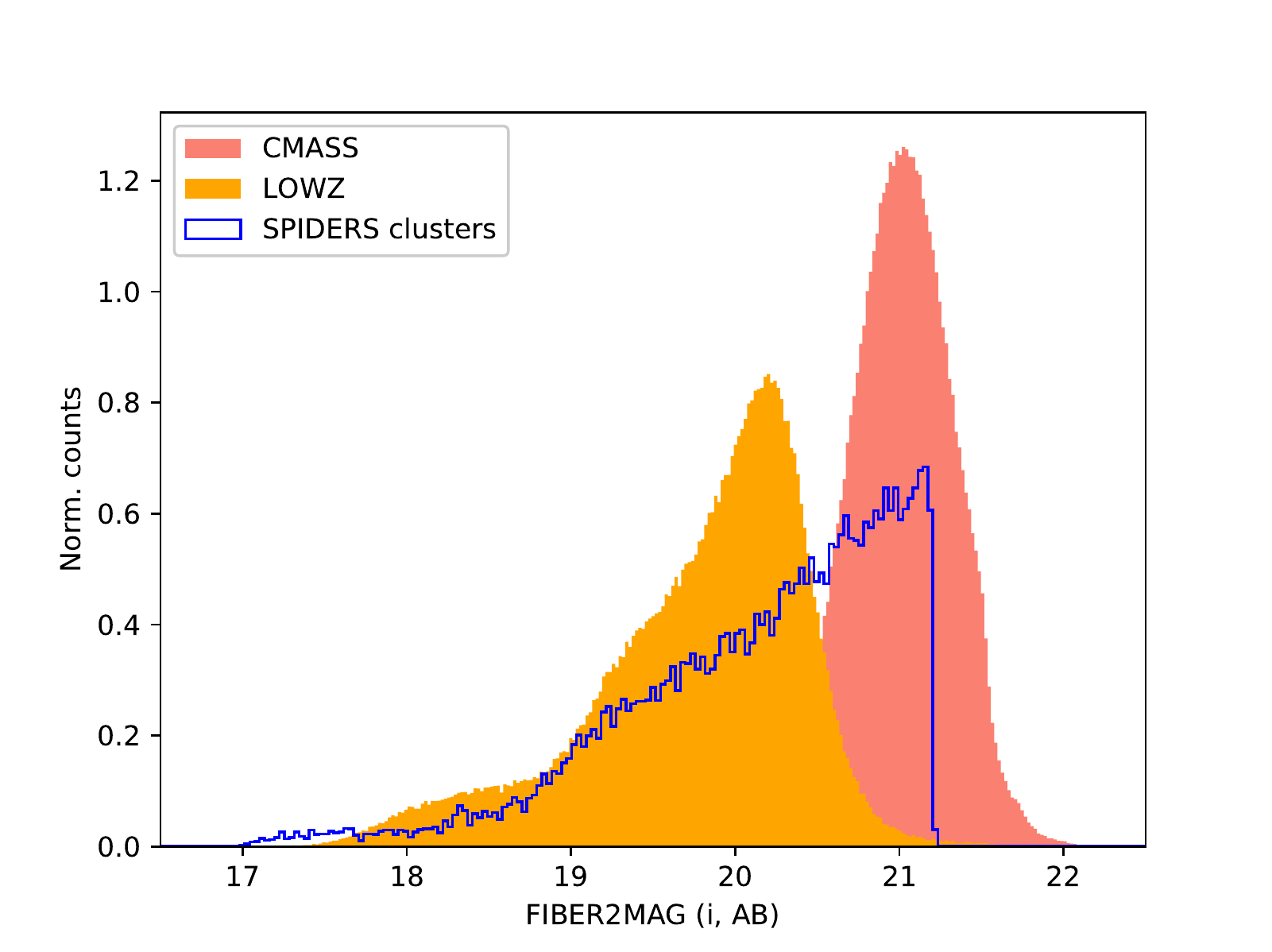}
    \caption{Distribution of the $2\arcsec$ fibre diameter magnitude (\texttt{FIBER2MAG}) in the SDSS-$i$ band for all SPIDERS targets with a reliable redshift. As a comparison, the distributions for the BOSS CMASS and BOSS LOWZ samples are also shown.}
    \label{fig:distrib_fiber2mag_i}
\end{figure}

\begin{figure}
	\includegraphics[width=\columnwidth]{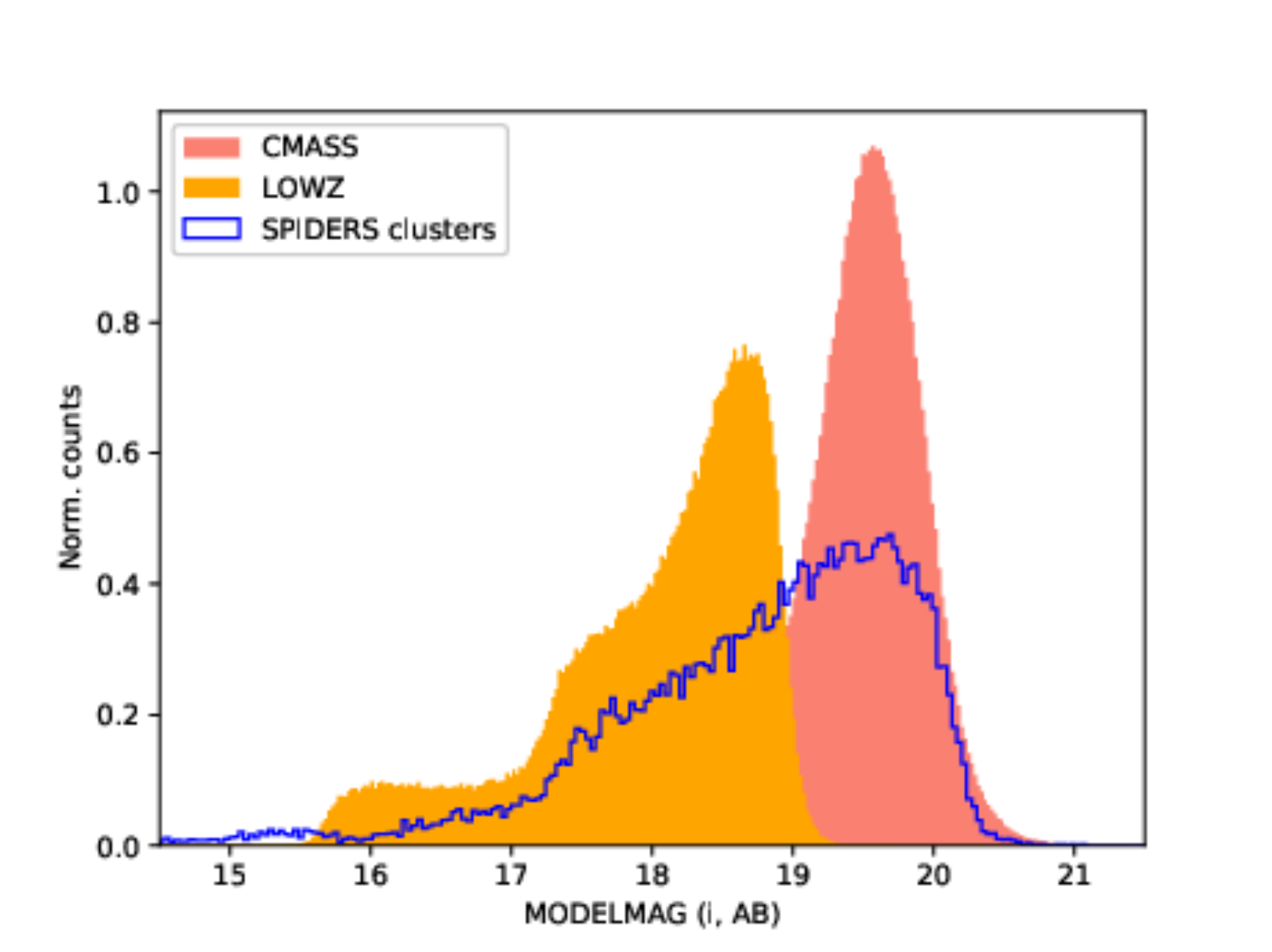}
    \caption{Same as Figure~\ref{fig:distrib_fiber2mag_i} but for model (total) magnitudes in the SDSS-$i$ band.}
    \label{fig:distrib_modelmag_i}
\end{figure}

Figure~\ref{fig:distrib_fiber2mag_i} shows the distribution of "fiber2mag" magnitudes in the SDSS-$i$ filter for SPIDERS and CMASS/LOWZ targets. These values scale directly with the flux entering a spectroscopic fibre. By design, it exhibits a sharp cut-off at $i=21.2$. The distribution is strongly skewed towards fainter sources, as expected from the shape of the luminosity function of cluster galaxies. The relative locations of this distribution compared to the CMASS and the LOWZ targets are consistent with the distributions shown on Figs.~\ref{fig:distrib_sn_median} and~\ref{fig:distrib_rchi2diff_noqso}. This behaviour is a direct consequence of the homogeneity of the spectral types (passive galaxies mainly) and consistent with the flux being the main driver of the spectral quality. Finally, Figure~\ref{fig:distrib_modelmag_i} represents the SDSS "model" magnitude distribution of targets, more representative estimate of the total light emitted by the galaxies in the $i$ filter. The smoother cut-off is located around $i_{\rm model} \simeq 20.2$.

\begin{figure}
	\includegraphics[width=\columnwidth]{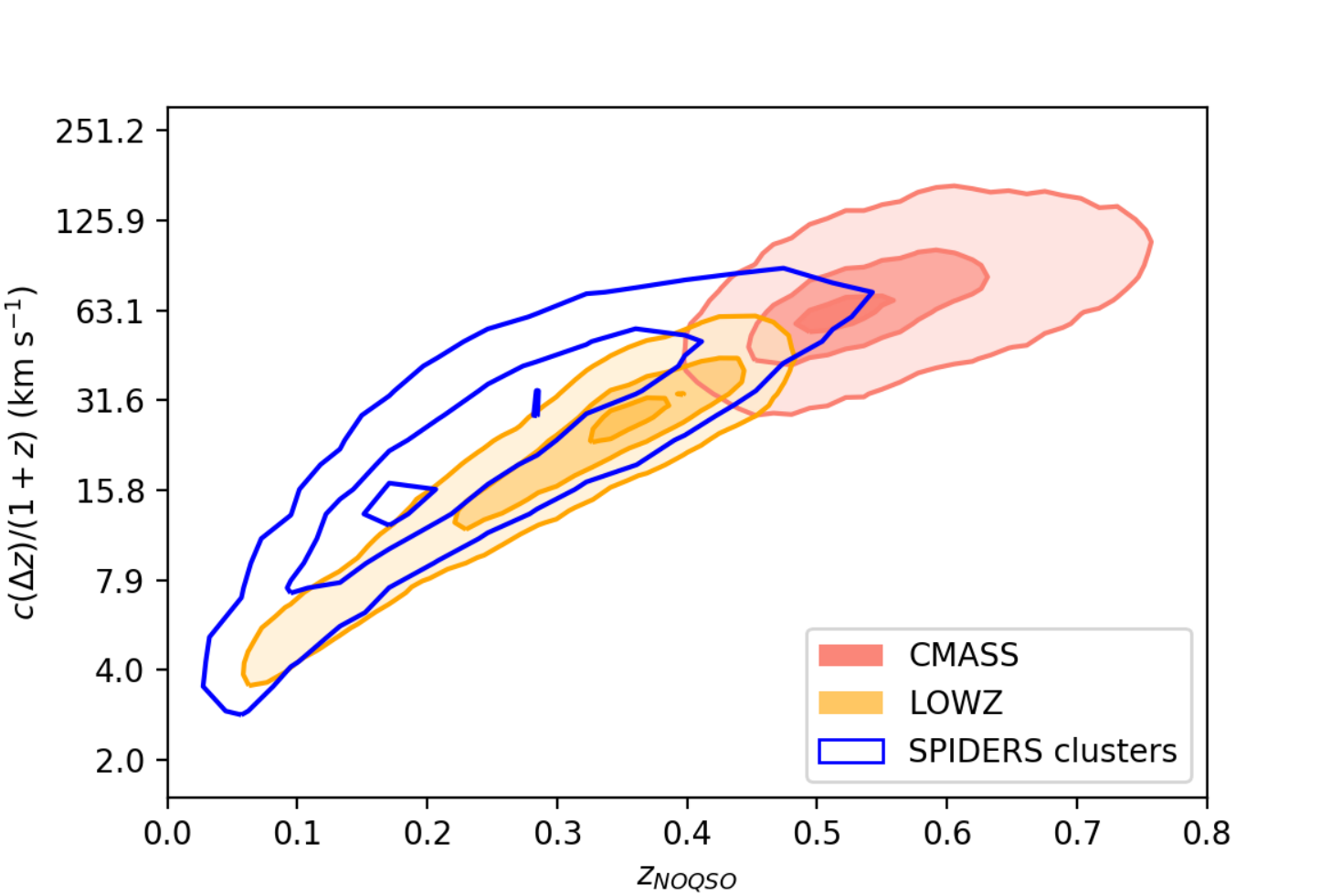}
    \caption{Distribution of best-fit redshift ($z_{\rm NOQSO}$) and redshift measurement uncertainty ($\Delta z$) for all SPIDERS targets with a reliable fit. The $y$-axis precisely shows the velocity uncertainty at the redshift of the target, $c\Delta z/(1+z)$.  As a comparison, the distribution for the BOSS CMASS and BOSS LOWZ samples are also displayed. Contours encompass 10\%, 50\% and 90\% of the total distributions.}
    \label{fig:distrib_z-z_err_noqso}
\end{figure}

SPIDERS galaxy cluster confirmation relies on "\_NOQSO" best-fitting values, which are obtained by discarding quasar (QSO) models from the spectral template library \citep{bolton2012}. In particular, object redshift ($z$) and associated uncertainties ($\Delta z$) are extracted from columns \texttt{Z\_NOQSO} and \texttt{Z\_ERR\_NOQSO}. Fig.~\ref{fig:distrib_z-z_err_noqso} presents the distribution of targets in the $z$-$\Delta z$ plane for the SPIDERS spectroscopic sample. On average, cluster targets reach $z \sim 0.6$ and their redshift distribution peaks at $z \sim 0.2-0.3$. The typical redshift uncertainties ($\Delta z \sim 0.2-1 \times 10^{-4}$) increase with redshift and are slightly higher than those of the LOWZ and CMASS samples. Again, this behaviour follows expectations from targeting satellite galaxies in massive halos. Following \citet{bolton2012} and \citet{dawson2016}, the pipeline uncertainties on individual redshift estimates should be multiplied by a factor of up to 1.34 in order to match tests done with repeat observations in BOSS. The median uncertainty for SPIDERS cluster redshifts with a reliable fit is approximately 20\,km\,s$^{-1}$. This value is well below the typical 100--1000\,km\,s$^{-1}$ galaxy velocity dispersions in galaxy clusters. Less than 1\% of the objects have uncertainties above 60\,km\,s$^{-1}$ and five objects show greater than 100\,km\,s$^{-1}$ uncertainties. Individual galaxy redshifts uncertainties are taken into account at the galaxy cluster catalogue production stage during visual inspection; extremely large values are usually interpreted as a redshift failure by inspectors and lead to discarding the corresponding object from the membership list \citepalias{kirkpatrick2020}.

Each successful redshift determination is associated with a spectral template. The vast majority of best-fit templates correspond to the "GALAXY" type of SDSS (as indicated by the \texttt{CLASS\_NOQSO} column in the DR16 spectroscopic sample). The 117 remaining spectra, among the 29277 reliably characterised ones, were fitted best with a "STAR" template at redshift zero. This relatively low fraction highlights the high-purity of the photometric classification and targeting algorithms.

The numbers presented in this section underscore the high overall spectral quality of the dataset acquired as part of the SPIDERS galaxy cluster program. The combined reliabilities of the eBOSS pipeline automated classification and redshift determination considerably ease the galaxy cluster construction process, since most of the individual galaxy spectra do not require further visual inspection. In addition, the spectroscopic success rate close to 100\% makes the computation of the final selection function a relatively simple task; i.e., the survey strategy allows one to shift the effort on the galaxy cluster membership and confirmation rather than on the validation of individual galaxy redshifts.

	\subsection{The SPIDERS reference galaxy cluster sample\label{sect:spicatalogue}}

Combining SPIDERS spectra with observations acquired during previous SDSS campaigns provides a large homogeneous database of redshifts whose prime usage is X-ray galaxy cluster confirmation. We briefly introduce here both cluster catalogues constructed from this combined dataset. Details of their construction, content, access and usage is provided in separate papers, cited in the following paragraphs.

		\subsubsection{Sampling and completeness of cluster red-sequences}

\begin{figure*}
	\includegraphics[width=\linewidth]{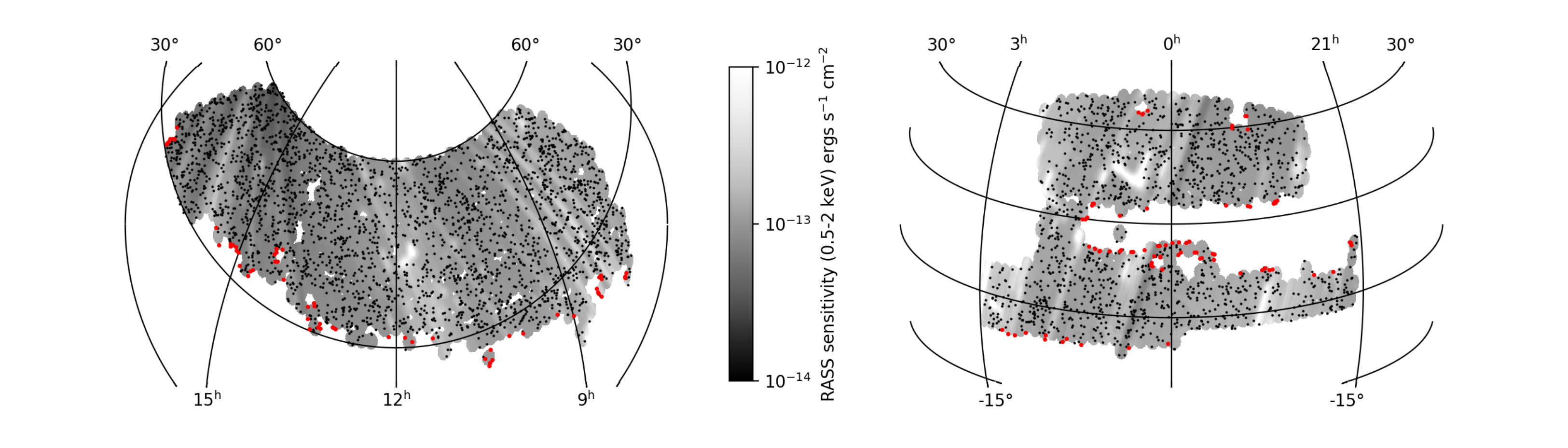}
    \caption{Distribution of CODEX cluster candidates (richness above 10) in the NGC (left) and SGC (right). Each point is located at the X-ray coordinates of a candidate. Red points indicate systems missing one or more redshifts in the list of assigned targets. The background shading represents the CODEX X-ray sensitivity, much as in Fig.~\ref{fig:sky_coverage}.}
    \label{fig:2d_completeness}
\end{figure*}

Figure~\ref{fig:2d_completeness} displays the sky location of individual cluster targets, with emphasis on spectroscopic completeness. Clusters missing one or several redshifts in their list of assigned targets are highlighted as red dots. These are mostly located at the border of the survey. At those positions, unobserved neighbouring spectroscopic plates can compensate for the missing data. The localised nature of incomplete systems and their low number is an outstanding result of the SPIDERS spectroscopic survey, enabling precise computation of the spectroscopic selection function of the final sample, thanks to the reproducible targeting process.
The spectroscopic completeness of X-CLASS galaxy cluster candidates follows a similar pattern on the sky. Only 11 systems lack one or more redshifts among the 142 located within the survey footprint.

\begin{figure*}
	\includegraphics[width=\linewidth]{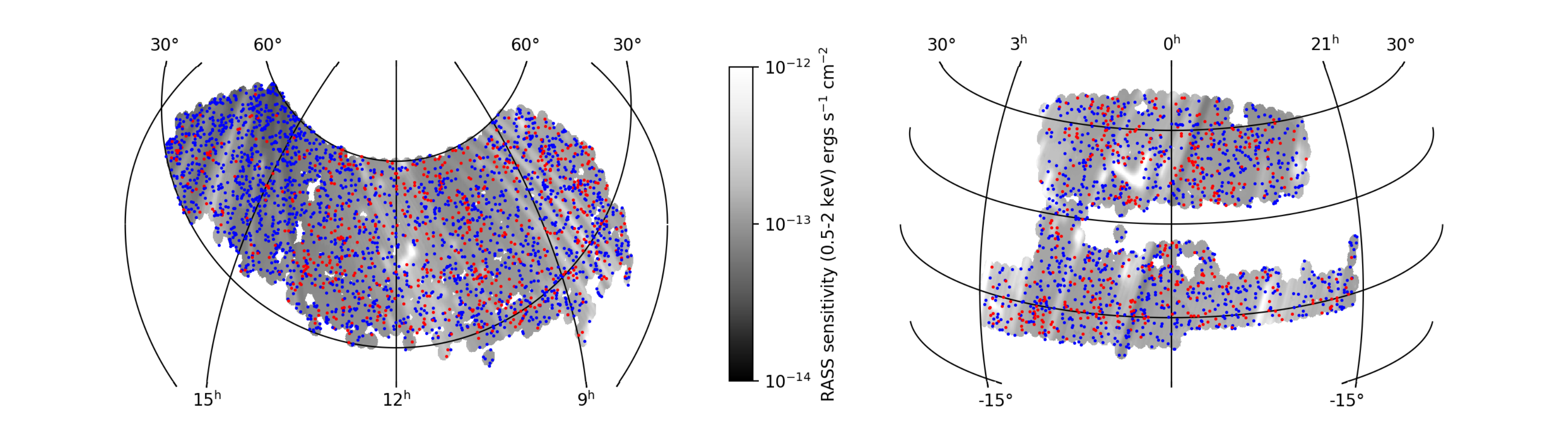}
    \caption{Same as figure as Fig.~\ref{fig:2d_completeness}, but now colour-coding reflects the number of spectroscopic redshifts collected in the red-sequence of each candidate ($p_{\rm mem} > 0.05$), regardless of whether they truly belong to the candidate system; clusters plotted in red have 15 or more measured redshifts.}
    \label{fig:2d_nhasz}
\end{figure*}

Figure~\ref{fig:2d_nhasz} similarly represents candidate clusters, now colour-coded by the number of spectroscopic redshifts collected for their red-sequence. Such targets may eventually not belong to the final set of confirmed clusters. The typical (median) number of spectroscopic redshifts per red-sequence is nine. In this figure, clusters with at least 15 galaxy redshifts from the red sequence are plotted in red. It is clear from this figure that the \texttt{eboss3} chunk, located at right ascensions between approximately 14h and 17h, significantly differs from the bulk of the survey. This result is a direct consequence of the targeting algorithm being adapted to the locally greater RASS sensitivities (Sect.~\ref{sect:target_selection}), which led to redistributing the fixed number of allocated spectroscopic fibres onto a larger pool of CODEX cluster candidates. Such a trend is absent from the X-CLASS targeting strategy, which shows a constant number of thirteen (median) spectroscopic redshifts per red-sequence.

		\subsubsection{The CODEX/SPIDERS catalogue}
		
\citet{kirkpatrick2020} \citepalias{kirkpatrick2020} describe the CODEX galaxy cluster sample validated with SPIDERS spectroscopy. This sample consists of 2,740 unique X-ray systems confirmed by combining an automated membership assignment performed by an algorithm and a series of refinements issued from collaborative visual screening performed by eleven experts. A moderation process combines individual inspections and eventually determines a unique membership list (and a unique cluster redshift, velocity dispersion, X-ray luminosity, mass, etc.) for each confirmed system. A final cleaning and merging is performed manually; this action allows removal of obvious duplicates and flags merger systems.
The redshift range of this sample is $0.016 < z < 0.677$ with a typical (statistical) uncertainty on each systemic redshift $\Delta_z/(1+z)= 6 \times 10^{-4}$. The median number of spectroscopic members per system is 10 (the mean is 12) and ranges between 3 and 75. The median number of visual inspectors per validated system is three, and a  minimum of two people validated a galaxy cluster. Scaling from their X-ray luminosities, the typical mass\footnote{Expressed here as $M_{200c}$, the mass within a radius $R_{200c}$. This is the radius enclosing an average density 200 times above the critical density at the cluster redshift.} of those systems amounts to $M_{200c} \simeq 4 \times 10^{14} M_{\odot}$, although individually precise mass values cannot be obtained with solely this dataset in hand. Excluding systems with less than 15 spectroscopic members, the mean velocity dispersion of galaxies within the SPIDERS clusters is about 620\,km\,s$^{-1}$.

\begin{figure*}
	\includegraphics[width=0.9\linewidth]{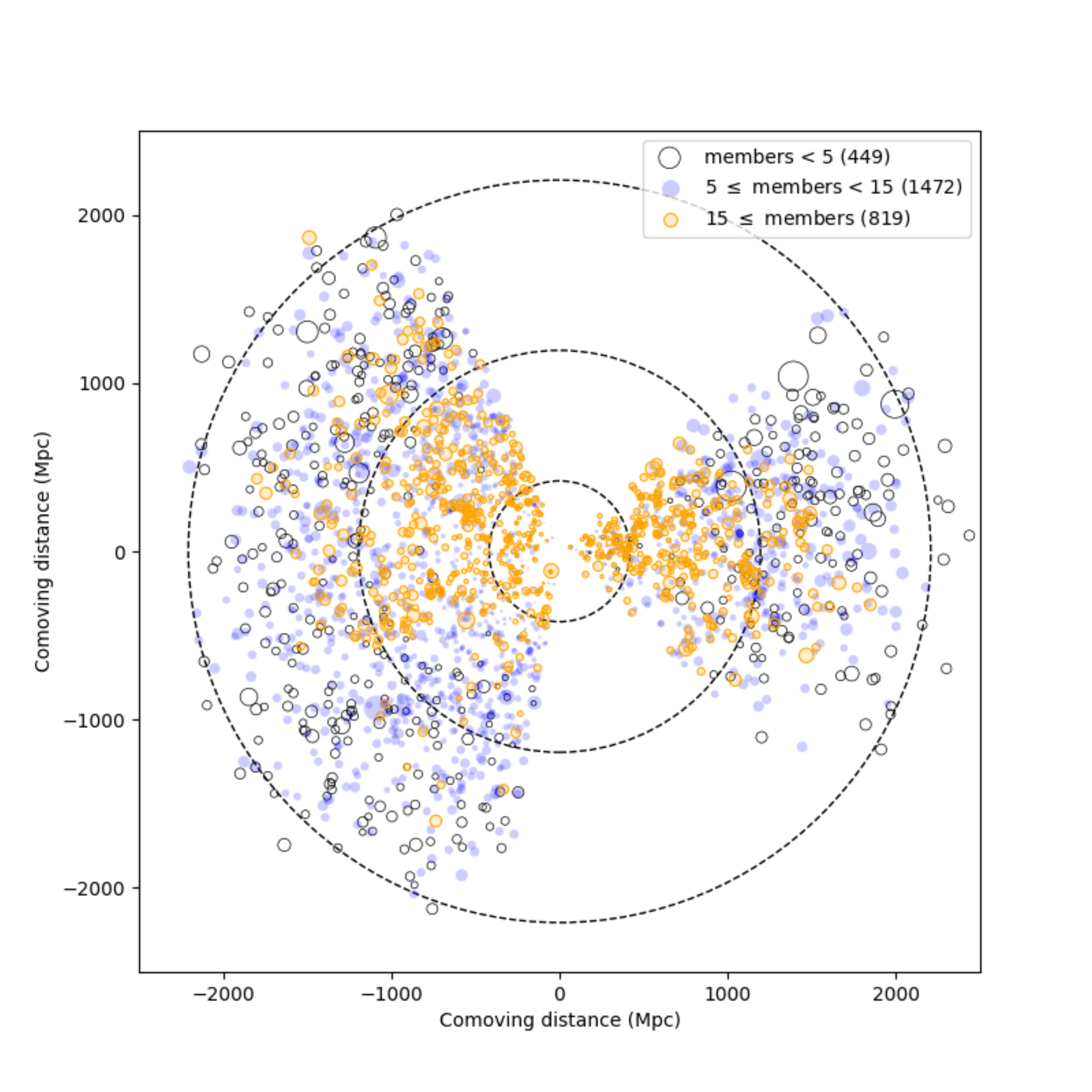}
    \caption{Representation of the CODEX spectroscopically validated catalogue of galaxy clusters (2,740 objects). The sample is fully described in \citet{kirkpatrick2020}. The observer is located at coordinates (0, 0) and comoving distances are relative to this position. Three-dimensional positions are collapsed along the axis perpendicular to the celestial equatorial plane, which is the plane of the figure. Positive $x$-axis values point towards R.A.=$0^{\circ}$; positive $y$-axis values point towards R.A.=$90^{\circ}$. Each circle represents a confirmed system, colour-coded by its number of spectroscopic members. The size of the circle for each cluster scales with the X-ray luminosity (hence, its mass). The selection function is well understood and the sample is not volume-limited, for instance Malmquist bias is clearly apparent in the increase of the typical luminosities of clusters with increasing redshift.}
    \label{fig:comoving}
\end{figure*}

Fig.~\ref{fig:comoving} provides an illustrative overview of the complete galaxy cluster catalogue. It represents each galaxy cluster by a circle whose size scales with its X-ray luminosity. The latter measurement may be impacted by various systematic uncertainties, among them contamination by AGN emission along the line of sight, possibly within the cluster itself. The locations of the systems are determined by their comoving distances from an observer located at (0, 0). The figure is drawn in the celestial equatorial plane. The left- and right-hand wedges correspond to the North and South Galactic caps, respectively. The colour coding indicates the number of spectroscopic members per system; the lower-left region of this plot comprises relatively less objects with $N_{\rm mem} > 15$. This area of the sky corresponds to the chunk \texttt{eboss3} where spectroscopic fibres (available in a limited amount) were evenly distributed among the more numerous X-ray sources. Malmquist bias clearly appears in the figure; at higher redshifts, the typical luminosity of the systems increases due to selection effects.

\begin{figure}
	\includegraphics[width=\columnwidth]{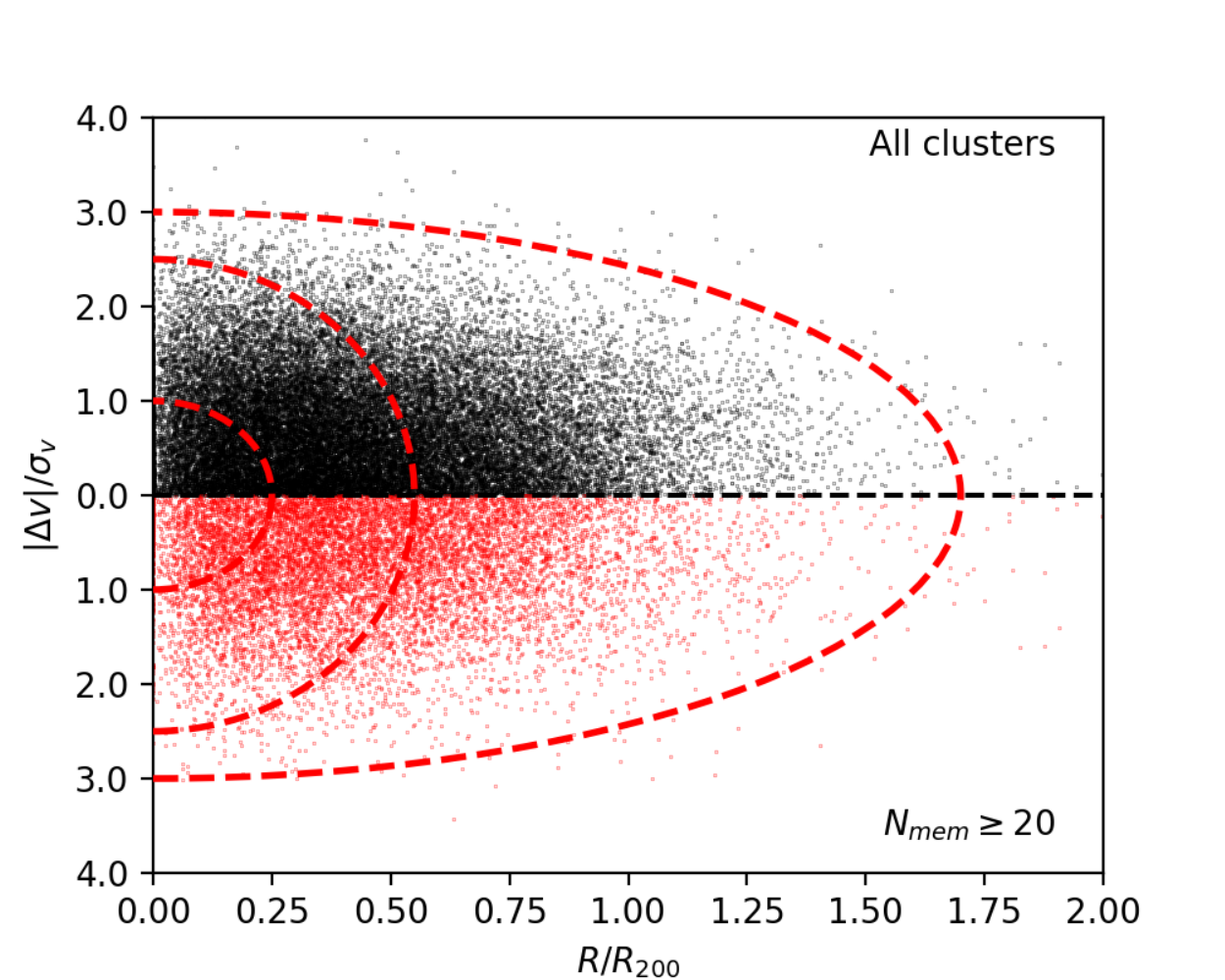}
    \caption{Representation of the SPIDERS red-sequence galaxies spectroscopically identified as cluster members (selected with condition \texttt{SCREEN\_ISMEMBER\_W} $\geq 0.75$) in the projected radius-velocity plane. Red points (the bottom panel) belong to clusters with more than 20 members, hence measured with high accuracy. The radius values are normalised by the projected $R_{200c}$ radius; velocity measurements are normalised by the velocity dispersion computed for each galaxy cluster.  Red dashed lines delineate bins designated `Inner', `Middle' and `Outer'; the external region is not labelled.}
    \label{fig:phase_space_bins}
\end{figure}

Thanks to the exquisite precision of each individual redshift measurement, it is possible to investigate the velocity dispersion of galaxies within the confirmed clusters. Figure~\ref{fig:phase_space_bins} displays the projected phase-space distribution of galaxies identified as cluster members in the CODEX sample. The projected distance is normalised to the value of $R_{200c}$ obtained from the relationship between X-ray luminosity and mass appropriate to these clusters. The velocity shift is normalised to the velocity dispersion measured from the velocity data itself, using the gapper or the biweight estimates \citepalias{kirkpatrick2020}. In the upper panel, all confirmed clusters are considered, thereby mixing badly determined velocity dispersion values with better measured ones. In the bottom panel, only those clusters whose number of spectroscopically identified members exceeds 20 are represented. These clusters benefit from high-quality systemic redshift measurements and robust determinations of the velocity dispersions of their member galaxies. The reliability of the $R_{200c}$ normalisation value mainly depends on the X-ray luminosity measurement itself and should also be considered with caution. Such distributions are key in determining dynamical mass estimates of clusters, as demonstrated in \citet{capasso2019, capasso2020} for a subset of the SPIDERS catalogue. In these works, dynamical masses are obtained by fitting mass models to phase-space distributions similar to those shown in Fig.~\ref{fig:phase_space_bins}, after additional cleaning based on richness and red-sequence spectroscopic coverage.

		\subsubsection{The X-CLASS/SPIDERS catalogue\label{sect:xclass_cata}}

The X-CLASS galaxy cluster sample validated with SPIDERS spectroscopy consists of 124 unique systems identified and confirmed following a similar methodology to that for the CODEX clusters. However, some notable differences occur in the visual screening process, namely: i)~only one inspector participated in the visual screening and ii)~X-ray data carried heavier weight in the decision process. These differences are direct consequences of the increased angular resolution and positional accuracy of the XMM detections with respect to RASS. It is straightforward to co-locate optical data and X-ray data based on the centroid and extent of X-ray sources. Visual inspection thus reduces to checking the 2d+1d (sky+redshift) consistency between X-ray, red-sequence, and spectroscopic data.

The median number of spectroscopic members per confirmed galaxy cluster is 10; the mean is 12. These numbers are similar to those derived for the CODEX clusters. The mean redshift is $z=0.31$ and ranges from 0.04 to 0.60. Optical richness ranges from $\lambda = 5$ to $\lambda = 180$, and a fraction of the sample overlaps the CODEX sample described in previous paragraph. Interestingly, all X-CLASS confirmed clusters are `C1' detections, thus providing enough photons on XMM detectors to enable good centroid and surface brightness profile measurements, as well as luminosity and temperature measurements in fixed apertures for all systems \citep[e.g.][]{adami2018}.

	\subsection{Spectral properties of the cluster member population\label{sect:spectral_stat}}

Figure~\ref{fig:stacked_spectra_2d} displays all individual spectra of galaxies identified as members of the CODEX sample, sorted by redshift. These objects were selected by imposing \texttt{SCREEN\_ISMEMBER\_W} greater than 0.75, meaning that more than 75\% of visual inspectors identified those objects as members of a galaxy cluster \citepalias{kirkpatrick2020} A fraction of the spectra lack coverage beyond {$\sim 9100$ \AA} since their origin is in SDSS-I/II datasets, with limited bandwidth in comparison to the BOSS/eBOSS surveys. The displayed fluxes are rescaled with the $2 \arcsec$ fibre fluxes in the $i$ band. 
This figure highlights a homogeneous spectral dataset of red, passive galaxies with notable absorption features such as Ca~H+K, G-band, \ion{Mg}{I}, NaD, etc. Barely visible in this figure are a few galaxies with emission lines such as [\ion{O}{III}] and H$\alpha$, included `by chance' or by targeting algorithms other than SPIDERS.

\begin{figure*}
	\includegraphics[width=\linewidth]{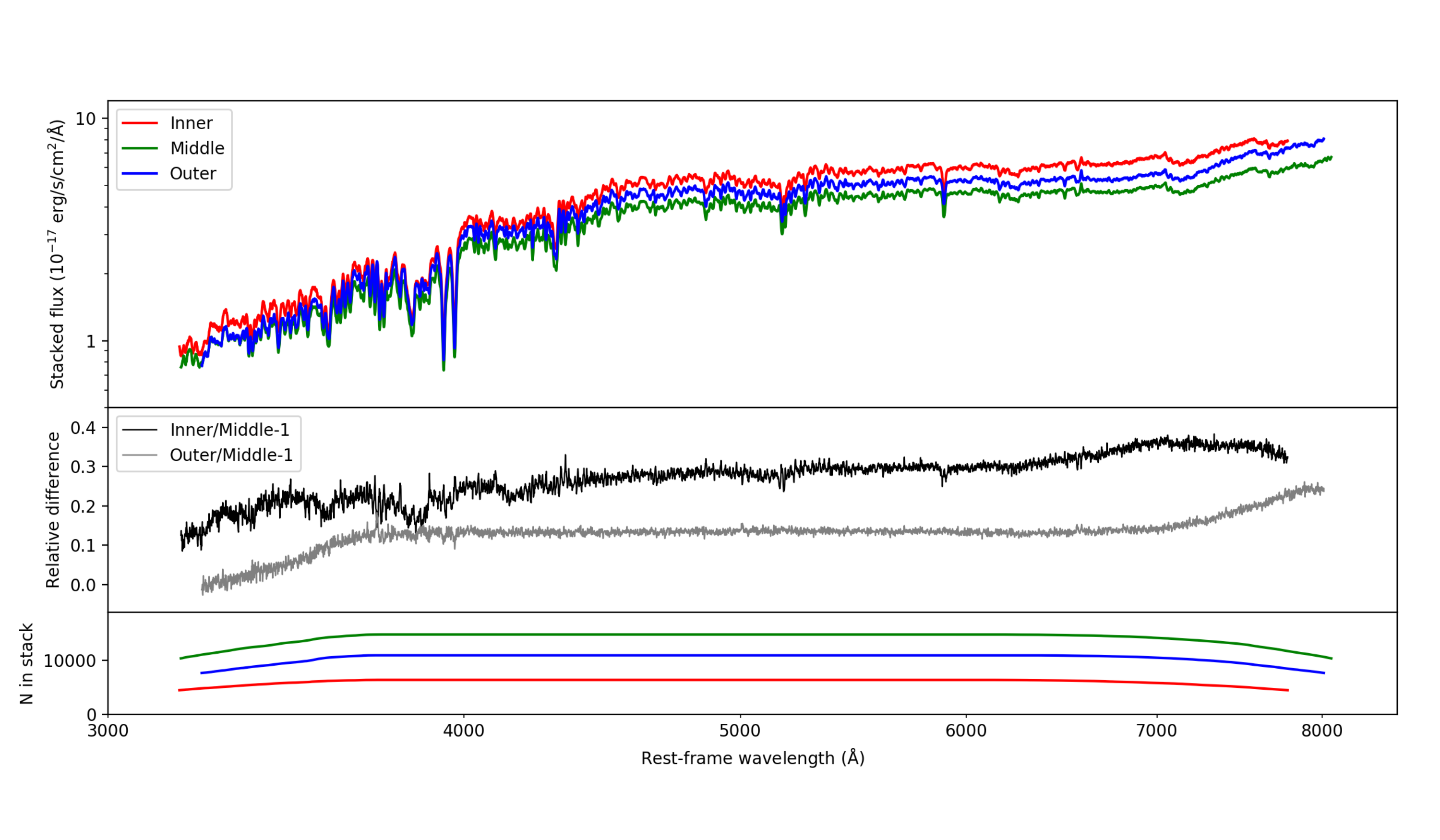}
    \caption{Median stacks of flux-calibrated spectra for all members within all SPIDERS confirmed galaxy clusters (top panel). Stacks are grouped into three bins drawn in $r/R_{200}-\Delta v/\sigma_v$ phase space (see text and Fig.~\ref{fig:phase_space_bins}). The three bins roughly correlate with galaxy infall time in clusters. The middle panel shows the relative difference between the `Inner' (`Outer') and the `Middle' stacks in dark (light) colours. The bottom panel indicates the number of spectra contributing to each wavelength for each composite. If a wavelength bin includes less than 70\% of the total number of spectra, it is discarded from this figure.}
    \label{fig:stacked_spectra_1d}
\end{figure*}

We fit stellar population models to stacks of SPIDERS red-sequence galaxies, following the empirical phase-space binning introduced in \citet{muzzin2014}. The bins are denoted `inner', `middle' and `outer' as drawn in Fig.~\ref{fig:phase_space_bins}. Stacks of the observed galaxy spectra are displayed in Fig.~\ref{fig:stacked_spectra_1d}. Although their definition is somewhat arbitrary, these bins are expected to correlate with time since infall onto the cluster \citep{muzzin2014}. The stellar population models (Chabrier IMF, MILES library) and fitting routine used allow one to decompose the observed continuum into a set of stellar populations of different ages \citep{comparat2017, wilkinson2017}. 
Regardless of the phase-space bin location, the stellar population is dominated by old and high metallicity populations. More than 90\% of the mass comes from stellar populations older than $10^{9.5}$ years. There is no hint of new star formation episodes. This result is consistent with \citet{bc03} models used for the red sequence selection in \citet{rykoff2014}.

	\subsection{Selection function of the confirmed sample and X-ray luminosity function\label{sect:sample_completeness}}

The set of galaxy clusters making the confirmed catalogue is a sample of the true underlying galaxy cluster population. The selection function is the statistical description of the link between both sets. Besides the geometrical selection on the sky, the sensitivity of the X-ray imaging to extended sources is the primary filter governing the selection. Then, the association to optical red-sequence galaxies enters as a secondary filter; it is a strong function of the richness and of the redshift of the galaxy concentrations. The spectroscopic follow-up, that is the chain of processes described in this paper, acts as an additional filter; it excludes systems with too few spectroscopic members. It is a complex function of the cluster properties, of the targeting algorithm and of the line of sight direction. Its computation is considerably simplified by taking advantage of the homogeneous and uniform coverage of SPIDERS. The spectroscopic selection is governed by the number of galaxies above a threshold luminosity, that is related to the cluster richness $\lambda_{\rm OPT}$, and by the distance of the object to the observer, that is given by its redshift $z_{\lambda}$.
\citetalias{kirkpatrick2020} found the spectroscopic completeness of the sample can be expressed as a step function: a red sequence of galaxies in the CODEX survey is spectroscopically confirmed by SPIDERS if $\lambda_{\rm OPT} \gtrsim 15 \times \zeta(z_{\lambda})$. Here $\zeta$ is the redshift-dependent scaling adopted by redMaPPer to account for galaxies brighter than $0.2 L^*$ and fainter than the SDSS magnitude limit \citep{rykoff2014}; it relates the raw galaxy counts to the richness estimate $\lambda_{\rm OPT}$ and the selection cut is easily interpreted in terms of the number of targetable galaxies in a cluster of a given richness $\lambda$. This number drops at $z \gtrsim 0.35$ as the SDSS limiting magnitude makes the fainter cluster galaxies undetectable. The factor $15$ encapsulates the effect of the target selection scheme, of fibre collisions, the photometric outlier rate, etc.
The cleaning criteria derived by \citet{finoguenov2020}, based on the 10\% RASS sensitivity, is a more conservative cut and reads : $\lambda > 22 (z/0.15)^{0.8}$ \citep[see also][and Fig.~\ref{fig:comoving_selec}]{iderchitham2020}.

\begin{figure}
	\includegraphics[width=\linewidth]{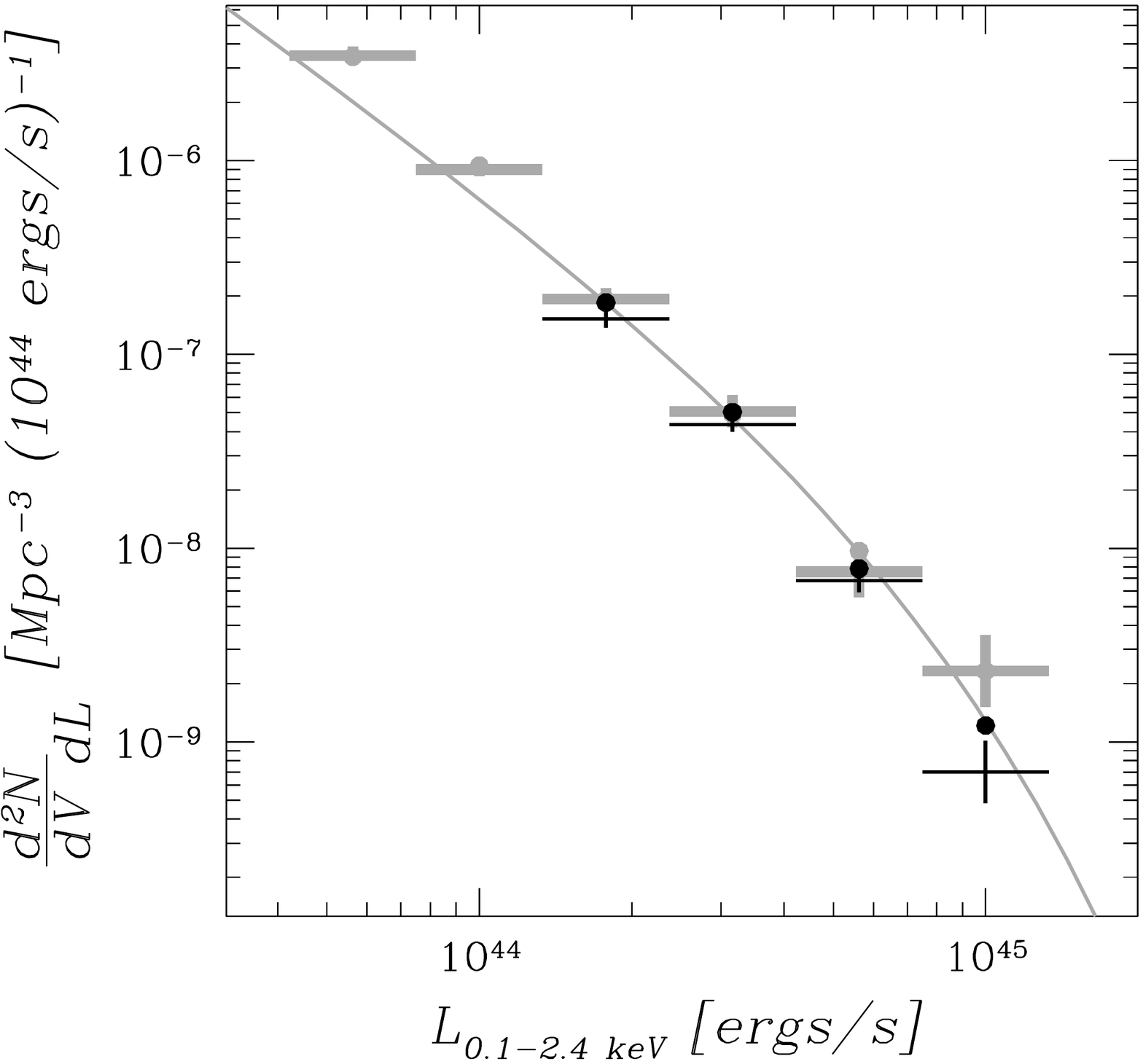}
    \caption{The X-ray luminosity function of the confirmed CODEX galaxy cluster sample in two redshift bins ($0.1<z<0.3$ in light grey and $0.3<z<0.6$ in black). Crosses are data points calculated using spectroscopic redshifts, these redshifts also serve in removing contaminating flux from AGN (see text); points represent data computed with photometric redshifts only, with no AGN removal. The line shows the Schechter function \citep{schechter1976} describing the REFLEX cluster luminosity function \citep{bohringer2002}.}
    \label{fig:xlf_model}
\end{figure}

This selection function is used in modelling the survey and catalogue creation. Fig.~\ref{fig:xlf_model} presents the X-ray luminosity function (XLF) derived from the CODEX galaxy cluster catalogue, using either spectroscopic redshifts or photometric (redMaPPer) redshifts. Only luminosity bins with completeness above 50\% are retained.
Calculations with spectroscopic redshifts (represented with crosses in the figure) incorporate removal of X-ray flux originating from AGN, stars and blazars, identified with the counterpart finding method `NWAY' \citep{salvato2018} in the entire SPIDERS analysis region \citep{comparat2020}. We only considered AGN with separations from the X-ray source smaller than $1.5\arcmin$ and in any case smaller than the separation between the X-ray peak and the optical cluster.

The X-ray luminosity function shown in Figure~\ref{fig:xlf_model} is not strongly impacted by the use of photometric redshifts (as expected) due to its slow evolution and the relatively large redshift bins used in this calculation. We found however that removal of spectroscopically identified AGN contribution has a significant impact on the shape of the XLF. While the effect remains moderate at low redshift ($z \lesssim 0.3$), at high redshift removal of AGN is essential. A 15\% shift in the XLF induces a shift on cosmological parameters comparable in magnitude with the current uncertainties on cosmological parameters.

\section{Discussion and conclusions}\label{sect:discussion}

	\subsection{A highly multiplexed spectroscopic follow-up programme}

SPIDERS represents an enormous investment of observing time, data processing, and management. It is, in several respects, unprecedented in the field of galaxy cluster studies, as witnessed by the following key numbers: close to 27,000 new redshifts were obtained in the project, distributed across more than 4,000 potentially massive halos within a volume approaching $5$\,Gpc$^3$ (comoving). Thanks to coherent data and information streams flowing from X-ray source detection down to final catalogues, its achievement is in large part due to the well-suited infrastructure of the SDSS-IV and eBOSS projects, including tools such as robust pipelines, uniform data bases, etc. This process is supported by significant team engagement and experience accumulated during years of operations.

The main objective of SPIDERS is to place cosmological constraints based on studies of the large-scale structure \citep[e.g.,][]{iderchitham2020}. Such studies typically require large number statistics, precisely measured observables, and wide ranges of halo mass and redshift, with vigilant control over selection effects. Therefore, the focus of this project was on obtaining redshifts for as many X-ray clusters as possible, ensuring reasonable balance between signal-to-noise ratio of redshift measurements (i.e.,~number of spectra per cluster) and coverage of the mass-redshift plane. Our targeting design accounts for the instrumental and survey limitations: fibre collision radius, spectrograph limiting magnitude, and wavelength coverage, etc. We have demonstrated that such a strategy is successful in obtaining a high-quality, uniform set of galaxy redshifts out to $z\sim0.6$ (Fig.~\ref{fig:stacked_spectra_2d}). Because such a set benefits from extremely reliable spectral classifications and redshift measurements, our design avoids inspection of individual spectra. Most of the manual workload is shifted onto verifying the cluster membership assessment preceded by an automated algorithm.

This visual inspection step plays a decisive role in the regime of low number of redshifts per cluster. Concerning SPIDERS/CODEX, all systems with optical richness above 10 have been visually inspected by at least two people \citepalias{kirkpatrick2020}. In total, more than 15,000 distinct visual inspections were collected.
Such a process is possibly impaired by biases inherent to human intervention. We believe, however, that chaining the automatic membership determination with two or more independent visual inspections does provide a robust determination, in line with most state-of-the-art X-ray cluster spectroscopic surveys \citep{guzzo2009, adami2018, sohn2018X}. Developments are in progress to further automatise the cluster membership association process (e.g.,~using machine learning). For instance, learning from the set of visual inspections sourced by SPIDERS, one may easily imagine an algorithm capable of associating spectroscopic redshifts with red-sequence galaxies and, ultimately, with X-ray sources, accounting for the delicate physics at play in the large-scale structure: mergers of various mass ratios, multiple projections along the line-of-sight, hot gas/galaxy centring shifts, measurement uncertainties, etc.

A key ingredient in such a project consists in a red-sequence identifier able to provide not only an estimate for cluster richness and redshift, but also a photometric membership probability for individual galaxies in a uniform manner across the entire survey area. This aspect was made possible in SPIDERS through running redMaPPer \citep{rykoff2014} over the $10,000\,\deg^2$ SDSS imaging area \citep{gunn1998, doi2010}. Indeed, identifying likely cluster members is crucial in guiding the selection of targets given a limited fibre$\cdot$hour budget. This aspect is particularly striking for the case of CODEX, where X-ray sources are extremely faint RASS sources with poor positional accuracies and barely existent morphological indices (centroid or size). The spectroscopy of X-CLASS sources is much more straightforward in this regard, since targeting and membership determination is informed by robust X-ray positions and morphologies.
Nevertheless, photometric red-sequence pre-selection has noticeable limitations: the definitive pool of acquired cluster members is by no means representative of the cluster galaxy population. It is biased towards the most luminous, most central and most `typically passive' galaxies (Fig.~\ref{fig:stacked_spectra_1d}). It also renders the sample selection function dependent on optical survey depth and peculiarities (e.g.,~masked areas, Fig.~\ref{fig:sky_coverage_imodel}). Such considerations translate into a technical challenge when computing correction effects, including Malmquist and Eddington biases \citep{finoguenov2020}. Ultimately, these complications can be alleviated by a well-considered usage of densely sampled fields \citep[e.g.][]{sohn2018RM, sohn2019COSMOS}, realistic numerical simulations \citep{comparat2020}, and refined selection cuts, making analysis samples sufficiently immune to various selection biases.

	\subsection{Outlook and prospects for SDSS-V/BHM and eROSITA}

SRG/eROSITA will repeatedly scan the entire sky and provide large numbers of targets for spectroscopic observation \citep{merloni2012, predehl2017}. Thanks to its X-ray quality, located mid-way between XMM-Newton and ROSAT, it is probable that targeting and data confirmation procedures will share similarities with the methods presented here in the context of SDSS-IV/SPIDERS. A number of improvements are foreseeable. These will take advantage of better resolved X-ray imaging, in particular, enabling precise colocation of the hot gas and galaxy (stellar) content of clusters out to large redshifts. Moreover, the increased depth of optical photometric datasets \citep{mcmahon2013, des2015, shanks2015, flewelling2016, ibata2017, dey2019, kuijken2019} combined with the use of modern algorithms (red-sequence finders coupling X-ray and optical data, bayesian counterpart finders, machine/deep-learning methods, etc.) will enable better determination of targets and more accurate membership assessment.

The eFEDS programme will link SDSS-IV to SDSS-V, and it will lead to the spectroscopic identification of galaxy clusters detected in the CalPV programme of eROSITA (12 spectroscopic plates). eFEDS will constitute the first opportunity to improve on the methods outlined in this paper. After completion of the first full-sky scan, SDSS-V \citep{kollmeier2017} will obtain spectra of the brightest X-ray sources in the Northern and Southern hemispheres. The spectra of the fainter sources will be obtained through the 4MOST instrument on the ESO/VISTA telescope \citep{merloni2019, finoguenov2019}. Designing work- and data-flows able to retain knowledge of selection steps and confirmation biases, together with enabling massive confirmation of the most relevant sources for cosmological studies is at the heart of such endeavours. The completed SDSS-IV/SPIDERS project presented in this paper has pioneered a novel way and is opening the door for the coming scientific revolution promised by eROSITA on SRG.


\section*{Acknowledgements}

The authors thank the referee for their comments which increased the quality of this paper.
The authors thank E.~Rykoff for providing the data of Fig.~\ref{fig:sky_coverage_imodel}.
Some of the results in this paper have been derived using the HEALPix \citep{gorski2005} package, the Matplotlib package \citep{hunter2007} and Astropy,\footnote{\url{http://www.astropy.org}} \citep{astropy:2013, astropy:2018}. 
This research has made use of "Aladin sky atlas" developed at CDS, Strasbourg Observatory, France \citep{bonnarel2000, boch2014}. It has made use of TOPCAT \citep{taylor2005}.

Based on observations obtained with XMM-Newton, an ESA science mission with instruments and contributions directly funded by ESA Member States and NASA.

This work was supported by CNES. AF, CK and JIC acknowledge travel support from FINCA.
NC acknowledges travel support from the Institut Francais de Finlande, the Embassy of France in Finland, the French Ministry of Higher Education, Research and Innovation, the Finnish Society of Sciences and Letters and the Finnish Academy of Science and Letters.
CK and NC acknowledge this research was supported by the DFG cluster of excellence `Origin and Structure of the Universe' (\url{www.universe-cluster.de}).
AS is supported by the ERC-StG `ClustersXCosmo' grant agreement 716762, and by the FARE-MIUR grant `ClustersXEuclid' R165SBKTMA.

Funding for the Sloan Digital Sky Survey IV has been provided by the Alfred P. Sloan Foundation, the U.S. Department of Energy Office of Science, and the Participating Institutions. SDSS-IV acknowledges
support and resources from the Center for High-Performance Computing at
the University of Utah. The SDSS web site is \url{www.sdss.org}.

SDSS-IV is managed by the Astrophysical Research Consortium for the 
Participating Institutions of the SDSS Collaboration including the 
Brazilian Participation Group, the Carnegie Institution for Science, 
Carnegie Mellon University, the Chilean Participation Group, the French Participation Group, Harvard-Smithsonian Center for Astrophysics, 
Instituto de Astrof\'isica de Canarias, The Johns Hopkins University, Kavli Institute for the Physics and Mathematics of the Universe (IPMU) / 
University of Tokyo, the Korean Participation Group, Lawrence Berkeley National Laboratory, 
Leibniz Institut f\"ur Astrophysik Potsdam (AIP),  
Max-Planck-Institut f\"ur Astronomie (MPIA Heidelberg), 
Max-Planck-Institut f\"ur Astrophysik (MPA Garching), 
Max-Planck-Institut f\"ur Extraterrestrische Physik (MPE), 
National Astronomical Observatories of China, New Mexico State University, 
New York University, University of Notre Dame, 
Observat\'ario Nacional / MCTI, The Ohio State University, 
Pennsylvania State University, Shanghai Astronomical Observatory, 
United Kingdom Participation Group,
Universidad Nacional Aut\'onoma de M\'exico, University of Arizona, 
University of Colorado Boulder, University of Oxford, University of Portsmouth, 
University of Utah, University of Virginia, University of Washington, University of Wisconsin, 
Vanderbilt University, and Yale University.

Funding for SDSS-III has been provided by the Alfred P. Sloan Foundation, the Participating Institutions, the National Science Foundation, and the U.S. Department of Energy Office of Science.

Funding for the SDSS and SDSS-II has been provided by the Alfred P. Sloan Foundation, the Participating Institutions, the National Science Foundation, the U.S. Department of Energy, the National Aeronautics and Space Administration, the Japanese Monbukagakusho, the Max Planck Society, and the Higher Education Funding Council for England.


\section*{Data availability}

The data underlying this article are available in the SDSS website at \url{https://www.sdss.org}.



\bibliographystyle{mnras}
\bibliography{myreferences} 


\appendix

\section{Two-dimensional stack of cluster member spectra}

\begin{figure*}
	\includegraphics[width=\linewidth]{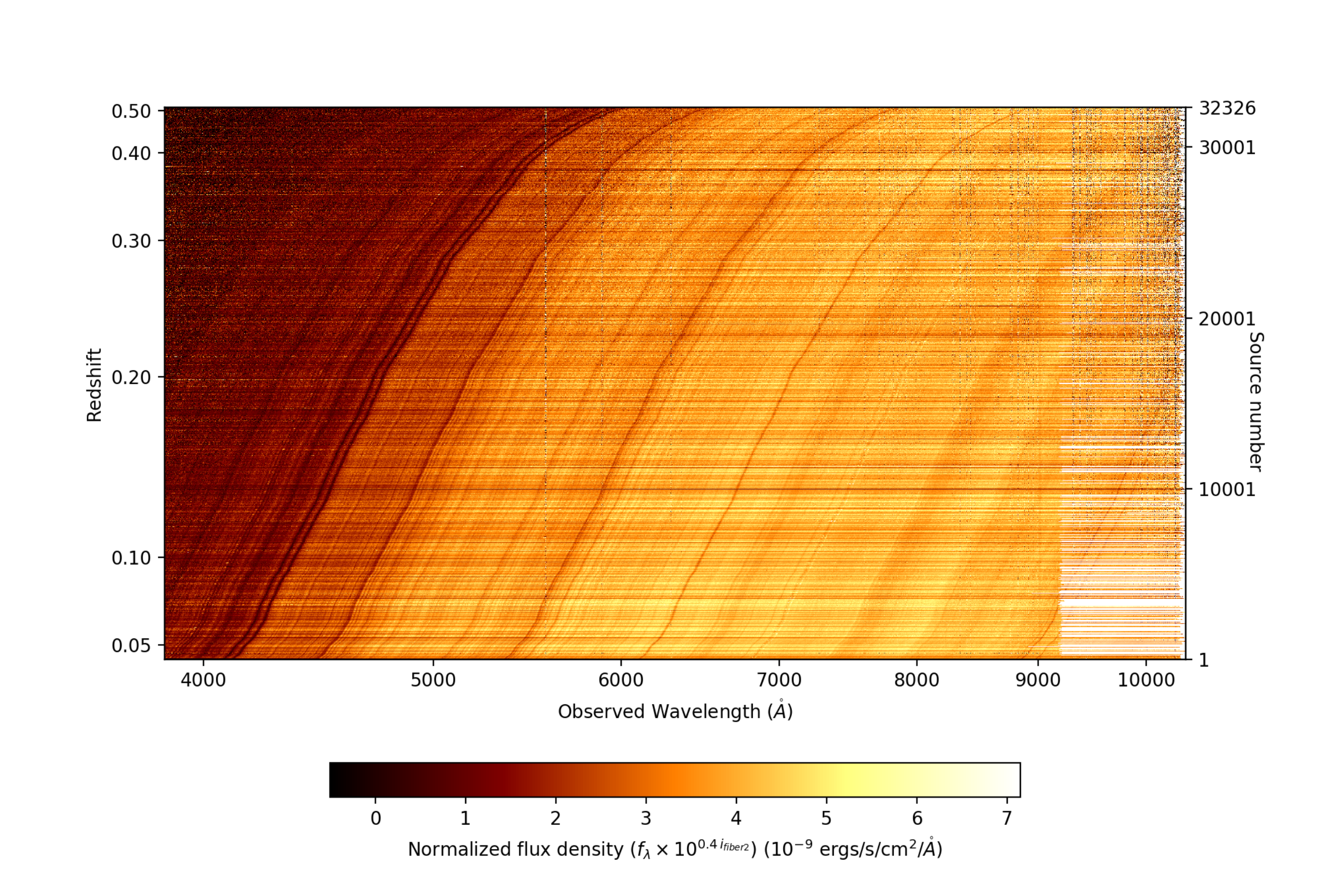}
    \caption{The flux-calibrated spectra corresponding to 32326 cluster members in the DR16 galaxy cluster catalogue (\texttt{SCREEN\_ISMEMBER\_W}$ \geq 0.75$). The colour linearly scales with the flux density in each wavelength pixel of a spectrum (units erg s$^{-1}$ cm$^{-2}$ \AA$^{-1}$), normalized to the fibre flux of each galaxy. Spectra are sorted according to their redshift (indicated on the $y$-axis). The observed wavelength scale is logarithmic. The main absorption features visible from $z = 0.013$ to $z = 0.679$ are \ion{Ca}{II}~H+K, G-band, \ion{Mg}{I}, NaD, etc. Some galaxies (barely visible here) display emission features, notably [\ion{O}{III}] and H$\alpha$. White stripes at observed wavelengths beyond {$9200$ \AA} represent missing data at wavelengths not covered by the SDSS-I/II spectrograph.}
    \label{fig:stacked_spectra_2d}
\end{figure*}

Fig.~\ref{fig:stacked_spectra_2d} illustrates the homogeneity of the sample of galaxies identified as members of SPIDERS clusters over a wide range of redshifts and masses.

\section{Sky distribution of "clean" CODEX clusters}

\begin{figure*}
	\includegraphics[width=0.9\linewidth]{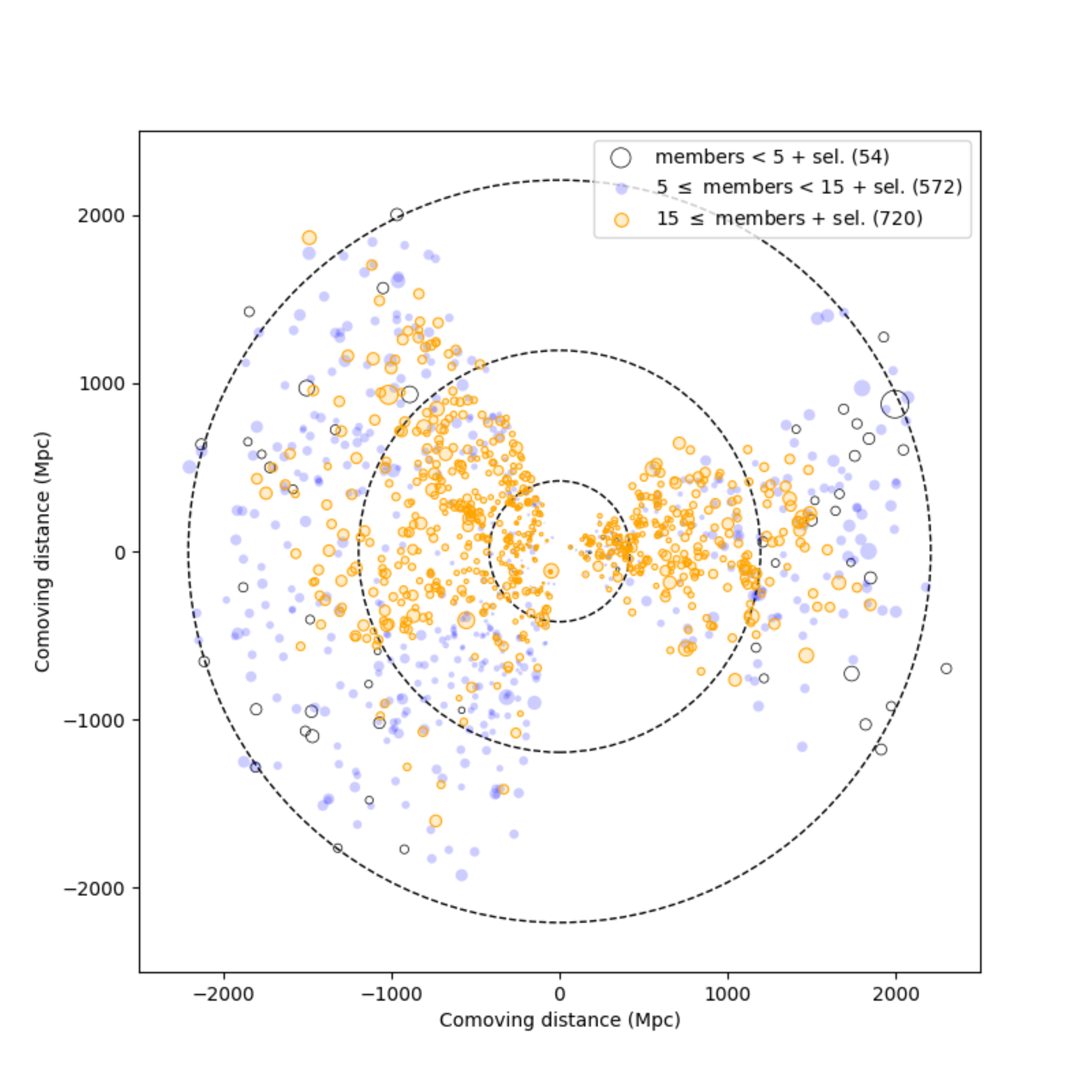}
    \caption{Same as Fig.~\ref{fig:comoving}, showing only the 1346 clusters passing the cleaning criteria of \citet{finoguenov2019} which derives from the 10\% RASS sensitivity and that reads: $\lambda > 22 (z/0.15)^{0.8}$. This selection happens to match well the low (5\%) contamination selection in \citet{klein2019}.}
    \label{fig:comoving_selec}
\end{figure*}

Figure~\ref{fig:comoving_selec} is a representation of all 1,346 clusters that pass the clean 10\% RASS sensitivity cut as defined in \citet{finoguenov2019}.

\bsp	
\label{lastpage}
\end{document}